# Click-Based Porous Cationic Polymers for Enhanced Carbon Diox-ide Capture

Alessandro Dani,*[a,b] Valentina Crocellà,[b] Claudio Magistris,[c] Valentina Santoro,[d] Jiayin Yuan,[a] Silvia Bordiga*[b]

[a.] *Department of Colloid Chemistry, Max Planck Institute of Colloids and Interfaces Am Mühlenberg 1 OT Golm, D-14476 Potsdam, Germany. E-mail: alessandro.dani@mpikg.mpg.de*
[b.] *Department of Chemistry, NIS and INSTM Reference Centre, University of Turin, Via Quarello 15, 10135 Torino, Italy. E-mail: silvia.bordiga@unito.it*
[c.] *Department of Chemistry and NIS Interdepartmental Centre, University of Turin, Via P. Giuria 7, 10125 Torino, Italy.*
[d.] *Department of Molecular Biotechnology and Health Science, University of Turin, Via Nizza 52, 10126 Torino, Italy.*

†Electronic Supplementary Information (ESI) available: materials syntheses, description of the techniques, HR-MS spectra and MSn fragmentation, additional nitrogen and carbon dioxide adosrption isotherms, thermogravimetric analyses data, SEM image and micro-calorimetry data. See DOI: 10.1039/x0xx00000x

**Abstract** Imidazolium based porous cationic polymers were synthesized using an innovative and facile approach, which takes advantage of the Debus-Radziszewski reaction to obtain meso-/microporous polymers following click-chemistry principles. In the obtained set of materials, click based-porous cationic polymers have the same cationic backbone whereas they bear the commonly used anions of imidazolium poly(ionic liquid)s. These materials show hierarchical porosity and good specific surface area. Furthermore, their chemical structure was extensively characterized using ATR-FTIR and SS-NMR spectroscopies, and HR-MS. These polymers show good performance towards carbon dioxide sorption, especially those possessing the acetate anion. This polymer can uptake 2 mmol/g of $CO_2$ at 1 bar and 273 K, a value which is among the highest recorded for imidazolium poly(ionic liquid)s. These polymers were also modified in order to introduce N-heterocyclic arbene along the backbone. Carbon dioxide loading for carbene polymer is in the same range of the non-modified versions, but the nature of the interaction is substantially different. Combined use of in-situ FTIR spectroscopy and micro-calorimetry evidenced a chemisorption phenomenon that brings to the formation of an imidazolium carboxylate zwitterion.

## Introduction

Porous organic polymers are a well-established among the field of porous materials.[1-5] More than ten years of research allowed for the discovery of many materials that differ in structure, porosity, functional groups crystallinity, and long-range order. Extensive research for tuning of these properties leads to target-specific porous polymers, useful to meet a variety of applications, such as gas or molecule storage,[6-8] separation, drug delivery,[9] electronics,[10, 11] and catalysis.[12, 13] Most of these materials are rich in phenyl and alkyne moieties,[14, 15] while only some of them possess ionic groups.[7, 16] One of the promising goals in the field of porous polymers is to introduce task-specific functional groups in the chemical structure of the porous network and target the topology and the structure of the material to a well-defined application.[1, 17] The synthetic pathways for these materials are usually time-demanding and require noble metal-based catalysts, since the common approaches involve Suzuki, Sonogashira-Hagihara cross-coupling and Yamamoto-type Ullmann cross-coupling reactions.[18] In recent years, "click-chemistry" has been exploited for the development of these materials, leading to advantageous and competitive syntheses in terms of time and costs.[19-23] A current challenge for porous polymers is to build up new structures, taking advantage of facile synthetic strategies, e.g. click chemistry-based ones, to transfer functional moieties into the polymers to define relevant industrial applications.

Among various kinds of porous polymers, poly(ionic liquid)s (PILs) are gaining more and more interest over the past years because of the high density of ionic liquid species in the macromolecular architecture, which lead to a broad range of applications.[24-29] Usually porous PILs are synthesized in a bottom-up approach via common radical polymerization, using cross-linker monomers and/or templates.[30-33] Some recent works describe the introduction of imidazolium or pyridinium ionic liquid functionalities inside a microporous polymer, basically by synthesis of different imidazolium or pyridinium functionalized monomers, which are respectively connected together by tetrahedral building unit, using different palladium based cross-coupling catalysis.[12, 13, 16, 34] Furthermore, two other groups reported about the closure of the imidazolium ring during the formation of the polymer network, in a two-step reaction. In this case the first step was the synthesis of the Schiff base polymer network; once the material was isolated, the imidazolium ring was closed in a second step.[7, 35]

In this work we describe how to obtain a set of "click-chemistry" based imidazolium porous poly(ionic liquid)s. These porous cationic polymers are obtained taking advantage of modified Debus-Radziszewski imidazolium synthesis, which follows the main principles of the "click-chemistry" defined by Kolb *et al.*[36, 37] The Debus-Radziszewski reaction for the production of imidazole dates back to more than a century ago,[38, 39] because of the high yield and the very mild conditions, this reaction remains the benchmark for industrial imidazole production.[40] A modified version of this reaction developed by Esposito *et al.* allows for the direct synthesis of imidazolium ionic liquids. This reaction already proved efficienct towards the synthesis of amino acid-derived imidazolium ionic liquids, linear poly(ionic liquid)s, and to cross-link polymers with dangling amino groups.[41-43] Its efficiency was also proven starting from aromatic amine and aldehyde as reactants.[44, 45]

Herein we exploited the Debus-Radziszewski imidazolium synthesis for the homocondensation of tetrakis(4-aminophenyl)methane, one of the most common tetrahedral building units used for the synthesis of microporous polymers. This click reaction leads to a network that expands in all the three dimensions, and the intrinsic steric hindrance of the monomers confers porosity to the resulting polymer. The reaction is irreversible because of the covalent closure of the thermodynamically stable imidazolium ring and runs under kinetic rather than thermodynamic control; therefore the resulting material is amorphous in nature. The as-obtained polymeric architecture faces the imidazolium cation linked through the position 1 and 3 to the main chain, differing from the common PILs that bear imidazolium as dangling groups. This imidazolium main chain polymeric architecture was described only in few works and showed increased thermal stability of the polymer and different chemical properties arising from the conjugation of the imidazolium with two phenyl rings.[46, 47]

Ionic liquids and poly(ionic liquid)s are well known materials for carbon dioxide adsorption.[48-50] We would like to take both the advantages of their intrinsic affinity towards this molecule and transfer it in microporous polymers, which are already known for their good gas adsorption properties.[51] The as-synthesized click reaction-based porous cationic polymers (from hereafter referred to as CB-PCPs) shows one of the highest carbon dioxide adsorption capacities ever reported among porous PILs, 2 mmol/g at 273K and 1 bar. CB-PCPs are imidazolium functionalized porous polymers, obtained via facile click-synthesis, and bearing all the common anions used in the field of PILs for carbon dioxide adsorption. N-heterocyclic carbenes (NHC) Are introduced in CB-PCPs, and the microporosity is tuned by varying the synthetic parameters. This set of CB-PCPs permit to understand how carbon dioxide adsorption properties are related to the structure and the porosity of the materials. In addition a deep physico-chemical study, performed with *in-situ* FTIR spectroscopy and adsorption micro-calorimetry, allows one to distinguish different interactions between carbon dioxide onto ionic polymers, or onto NHC bearing polymers.

## Syntheses

Click reaction-based porous cationic polymers (CB-PCPs) were synthesized according to a procedure reported in Scheme 1. Briefly, tetrakis(4-aminophenyl)methane was dissolved in a mixture of water and acetic acid. In a second vial formaldehyde and methyl glyoxal were dissolved in water and acetic acid. These two solutions were mixed together and the reaction proceeded immediately, as evidenced by solution color variation from yellow to dark brown, due to the development of the conjugated aromatic system; then the solution was heated at 80 °C for 12 hours (Scheme 1 part a) to complete the reaction. The acetic acid is used as catalyst and also ended up as acetate anion in the final product named CB-PCP **1.** The proposed mechanism of modified Debus-Radziszewski imidazolium synthesis is repored in Figure S1. Others CB-PCPs having different anions are obtained by anion exchange of CB-PCP **1** polymer solution with different salts. We used sodium tetrafluoroborate, bis(trifluoromethane)sulfonimide lithium salt, potassium hexafluorophosphate, and sodium trifluoromethanesulfonate to obtain respectively CB-PCP **2**, **3**, **4** and **5** (Scheme 1, part b). NHC was introduced by reacting polymer CB-PCP **1** with potassium *tert*-butoxide in anhydrous THF for 72 hours; the obtained sample is named CB-PCP **6** (Scheme 1, part c). All the polymers were purified according to their physical aggregation status: if in solution, they were dialyzed against water; if as a gel they were washed with water. Finally, powder CB-PCPs **1**-**5** were obtained by freeze-drying, whereas moisture-sensitive CB-PCPs **6** was washed with anhydrous solvent and freeze-dried from 1,4-dioxane. Detailed description of syntheses can be found in supporting information. A successful anion exchange offers an indirect evidence of successful closure of the imidazolium ring during the synthesis. In fact, only a cationic polymer is able to permanently bind various anions along the structure, without their leaching during the washing step. This new click-based synthetic approach makes PCPs and porous PILs even more green and sustainable.[19] The main strengths of this reaction are the following:

- It gives polymers in high yield (almost 100%).
- It only forms water as sub-product.
- It has a high thermodynamic driving force arising from the closure of the imidazolium ring.
- Reaction products are stable and the reaction is not reversible.
- 85% Efficiency in atomic economy is obtained.[52]
- Reaction takes place in very mild conditions (80 °C for 12 hours).
- Reagents are easily available except for the tetra-amine, which requires some synthetic steps.
- Reaction uses water and acetic acid as solvents.
- Final product is a solid (CB-PCPs **1a**, **1b**, **2-5**) and it is easily purified from the residual monomers by washing with water.

The modifeied Debus-Radziszewski reaction itself works at room temeperature and is rather fast (about 1 hour) when performed at molecoular level or in case of linear polymers.[41, 42] While in case of cross-linked polymer it is necessary to increase temperature and reaction time because of the increased viscosity.

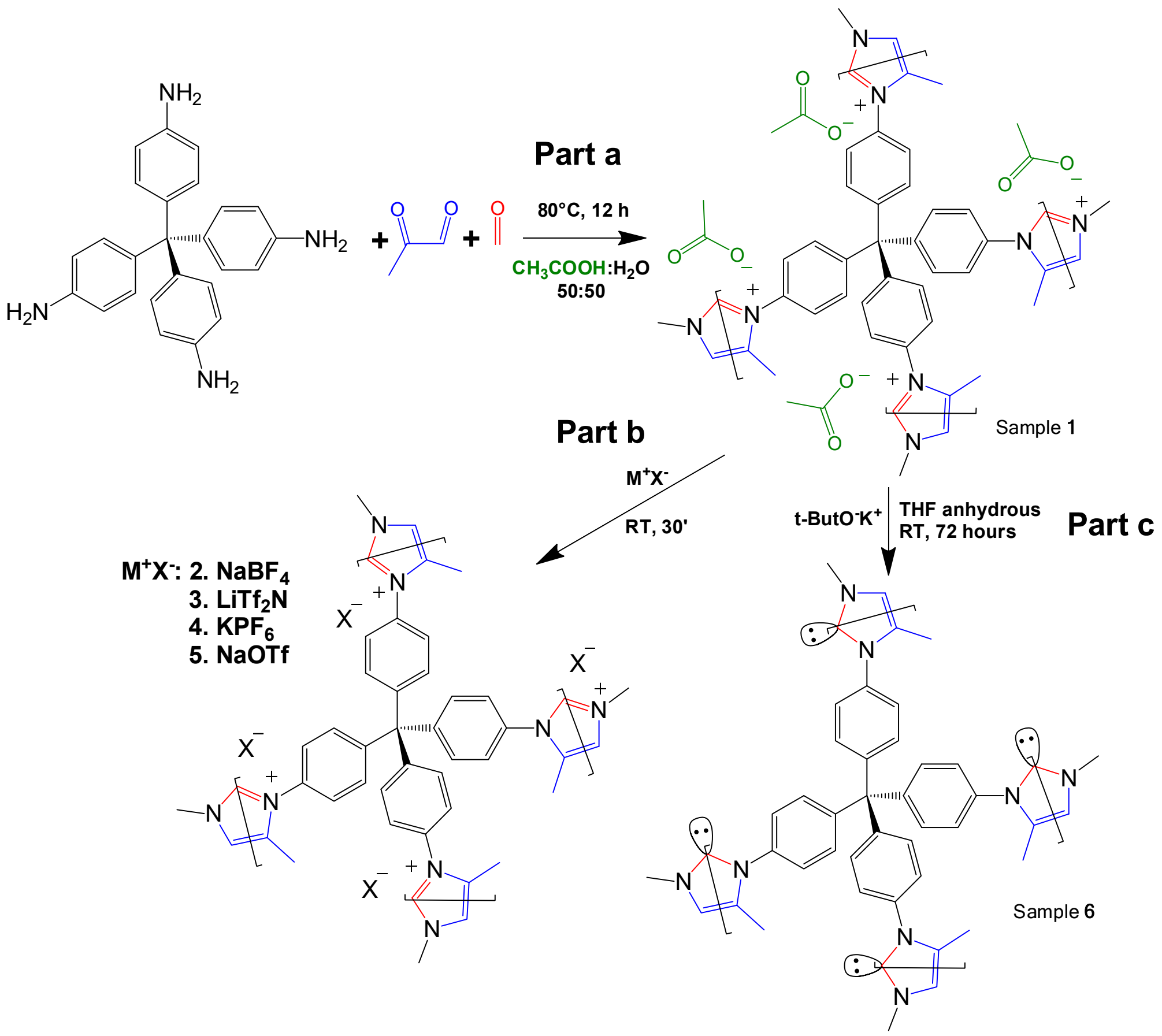

**Scheme 1**. **Part a:** the synthetic route towards the sample CB-PCP 1 starting from thetetrakis(4-aminophenyl)methane in a one-step reaction. **Part b:** the anion exchange towards the samples CB-PCP **2**, **3**, **4** and **5**. **Part c:** the synthetic route towards sample CB-PCP 6 introducing NHC from the imidazolium ring.

## Vibrational properties of CB-PCPs

The attenuated total reflection-infrared (ATR-IR) spectrum of sample **1**, reported in Figure 1 curve b, was collected as first evidence of the successful polymerization. The spectrum is characterized by a series of complex overlapped absorptions. In particular, the two bands located at 1504 and 811 $cm^{-1}$ can be ascribed to the ν(C=C) stretching of the conjugated aromatic system and to the τ(C-H) out of plane bending respectively.[53] These bands are also well evident in the spectrum of tetrakis(4-aminophenyl)methane monomer, reported as reference in Figure 1 curve a. The first, clear evidence of the occurred polymerization is the absence, in the spectrum of the CB-PCP, of the two peaked signals at 3156 and 3395 $cm^{-1}$ related to the ν(N-H) stretching modes of the amino groups of the monomer. In fact, in the 3500-3000 $cm^{-1}$ region, only a broad and ill-defined band is present, ascribable to the ν(O-H) stretching vibrations of physisorbed water. Unfortunately, the peculiar absorption band of the imidazolium ring, at around 1160 $cm^{-1}$, appears too weak to be detectable, due to the presence of the broad set of signals in the 1500-800 $cm^{-1}$ spectral range. For this reason, this specific absorption cannot be used as evidence of the successful polymerization. The band at 1705 $cm^{-1}$ is ascribable to the ν(C=O) stretching mode of the amide moiety of the chain terminals, generated by the reaction between the amino groups of the monomer and the excess of acetic acid. This chemical behavior is also well evident in the mass spectra of the dimer reported herein later. Both the symmetric and antisymmetric stretching modes of the acetate anion can be observed in the spectrum of sample **1,** at 1409 and 1566 $cm^{-1}$ respectively.[54] These bands are well evident also in the spectrum of sodium acetate reported as reference in Figure 1 curve c.

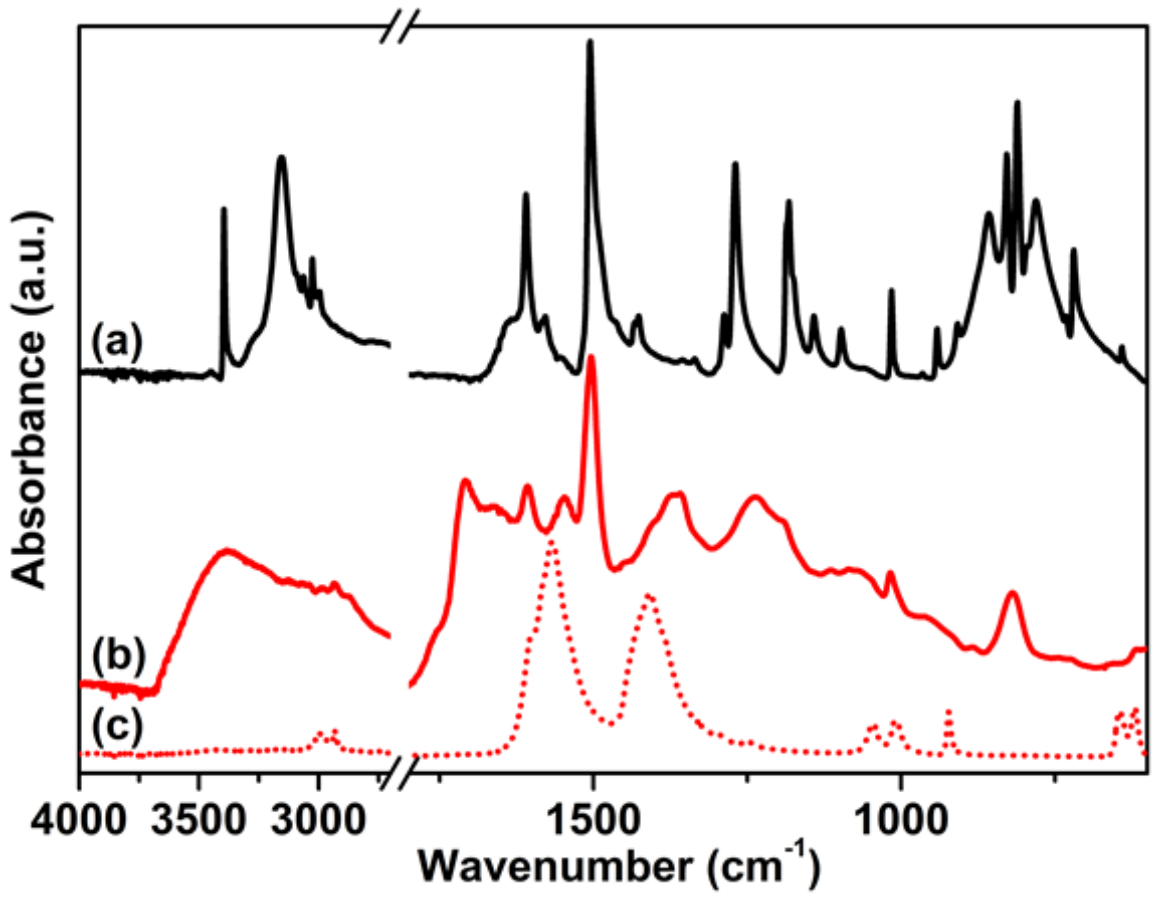

**Figure 1**. ATR-IR spectra collected in air of tetrakis(4-aminophenyl)methane (a), sample **1** (b) and sodium acetate (c).

Anion exchange with various salts was performed as previously described in order to replace the acetate anion and to obtain a set of different CB-PCPs. The ATR-IR spectra of all these materials were reported in Figure 2 and compared with the spectra of the corresponding salts. In all cases, the vibrational modes of the polymeric network are clearly overlapped with the spectral features of the employed anion. In particular, the following can be observed:

i. The ATR-IR spectrum of sample **2** (Figure 2 curve a) displays a medium broad band at 1060 $cm^{-1}$, assigned to the asymmetric vibrational modes of $BF_4^-$ anion,[55] and a signal at around 3152 $cm^{-1}$, overlapped with the broad band of physisorbed water, due to the combination of the imidazolium ring stretching mode and the $BF_4^-$ stretching vibrations.[56, 57] The reference spectrum of $NaBF_4$ (Figure 2 curve b) exhibits a single band at 1005 $cm^{-1}$ related to the $BF_4^-$ vibrational modes, this signal is red-shifted with respect to sample **2** due to the strong force field of $Na^+$ cation interacting with $BF_4^-$.

ii. In the case of sample **3** the ATR-IR spectrum (Figure 2 curve c) exhibits all the $Tf_2N^-$ ion peculiar vibrational modes (see asterisks) at 747 $cm^{-1}$ (ν(S-N)), 798 $cm^{-1}$ ((ν(C-S)+ν(S-N)), 1188 $cm^{-1}$ (ν($CF_3$)) and 1356 $cm^{-1}$ (ν($SO_3$))[58], overlapped with that of the polymer network, as highlighted by the comparison with the reference $LiTf_2N$ spectrum (Figure 2 curve d).

iii. A strong band at 841 $cm^{-1}$, due to the ν(P-F) stretching mode of the $PF_6^-$ anion, is evident in the ATR-IR spectrum of sample **4** (Figure 2 curve e). In the $KPF_6$ reference spectrum (Figure 2 curve f), this band is red-shifted to 805 $cm^{-1}$, because of the interaction between the ion pairs. In fact, the strong force field of the small $K^+$ cation affects the length of the P-F bonds of the $PF_6^-$ anion.

iv. Sample **5** presents, in its ATR-IR spectrum (Figure 2 curve g), three new intense bands at 1245 $cm^{-1}$, 1153 and 1027 $cm^{-1}$ ascribable to the ν($CF_3$) and ν($SO_3$) modes characteristic of $TfO^-$ anion, as referenced by NaOTf in Figure 2 curve h. [59]

The NHC formation in Sample **6** is evidenced by the appearance of a new strong adsorption band at 1605 $cm^{-1}$, arising from the stretching of the NHC ring (Figure 2 curve i).[7]

## Elemental analysis

The elemental analysis of sample **1** was performed in order to obtain quantitative information on the elemental structure of the material. The percentage of C, H and N, reported in Table S1, are respectively 74.78%, 5.55% and 9.11%. Those values are in perfect agreement with the calculated ones, pointing out a polymer structure having an elemental compostion corresponding to the suggested structure.

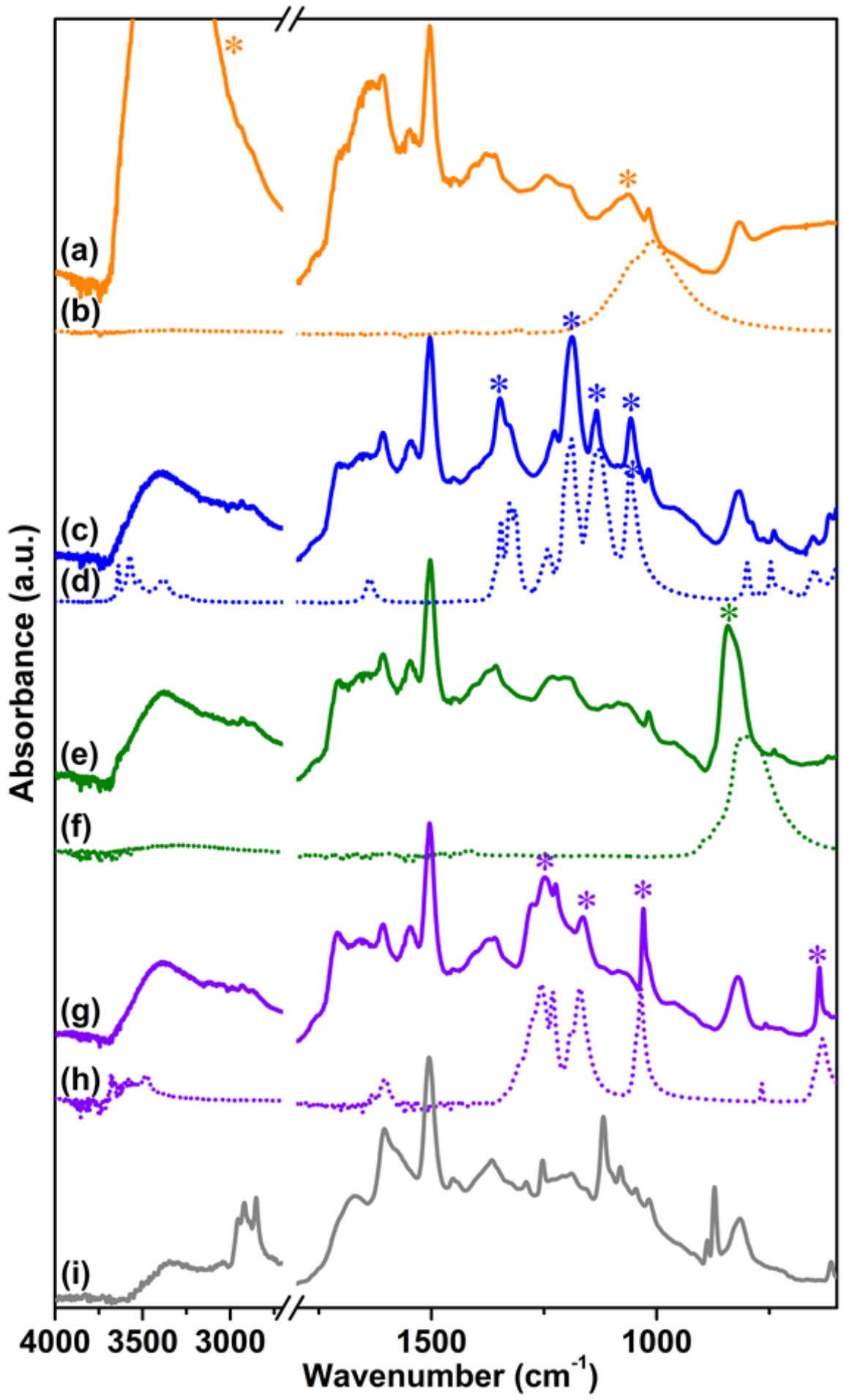

**Figure 2.** ATR-IR spectra collected in air of sample **2** (a), $NaBF_4$ (b), sample **3** (c), $LiTf_2N$ (d), sample **4** (e), $KPF_6$ (f), sample **5** (g), NaOTf (h) and sample **6** (i).

## Solid state NMR

In order to prove the structure of CB-PCP **1** and the presence of the imidazolium inside the polymeric network, $^1H$ (Magic Angle Spinning) MAS and $^{13}C$ Cross Polarization Magic Angle Spinning (CP-MAS) solid state NMR spectroscopy measurements were performed. Figure 3a shows the $^1H$ MAS solid state NMR of sample **1**, where two families of protons are clearly distinguishable. The one at lower chemical shift is related to the $–CH_3$ protons of the dangling methyl group in position 4 on the imidazolium ring and also to the $–CH_3$ protons of the acetate anion, whereas the other one is related to the aromatic protons along the polymeric backbone. Figure 3b and c report respectively the $^{13}C$ CPMAS solid state NMR spectra and a polymer scheme with the corresponding assignations. In the range between 100 and 150 ppm, there are the carbons constituting the phenyl and the imidazolium rings.

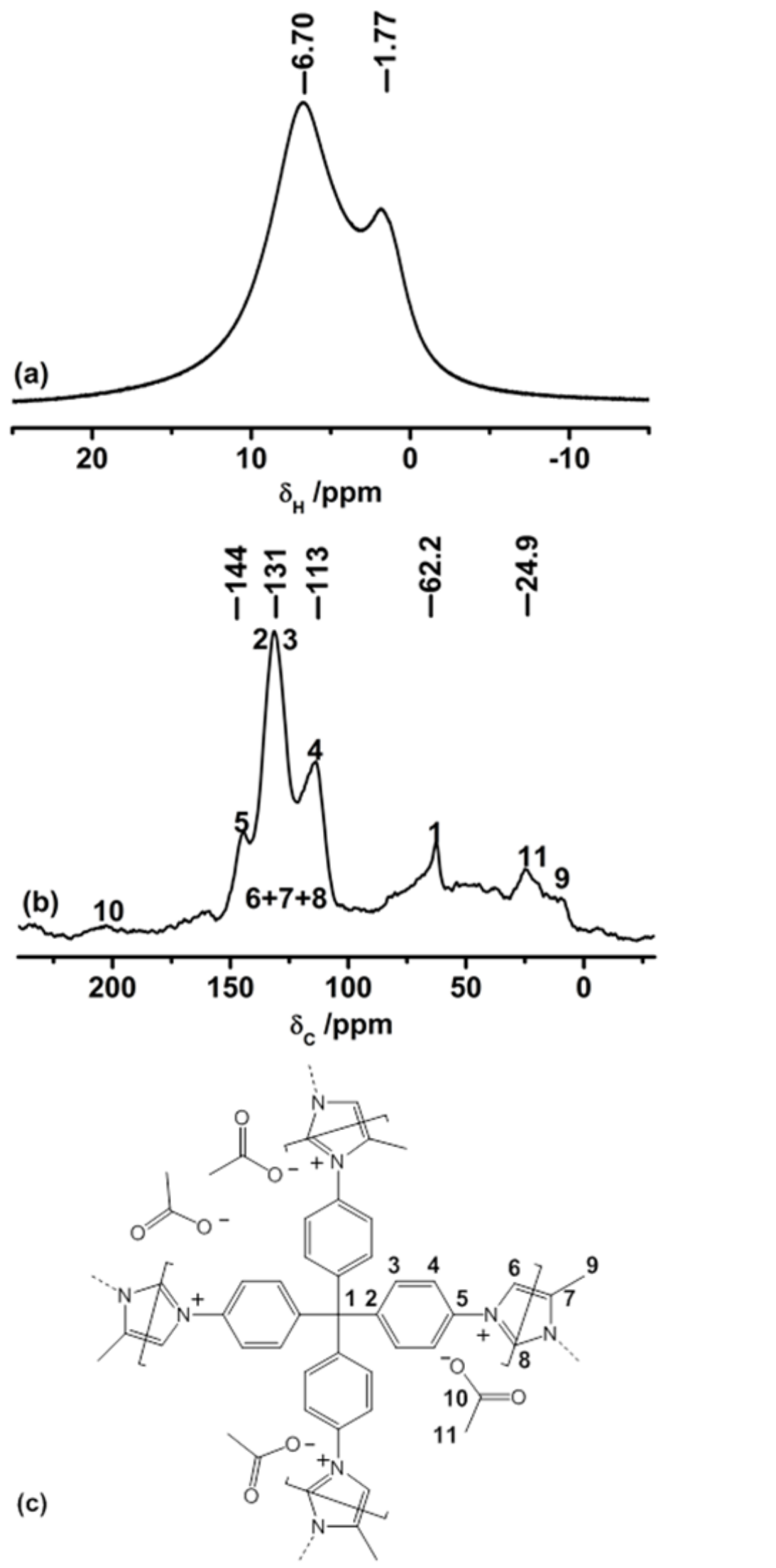

**Figure 3.** $^{1}H$ MAS (a) and $^{13}C$ CP-MAS (b) solid state NMR spectra of sample **1**. (c) sketch of sample **1** that evidence the assignment of $^{13}C$ CP-MAS spectrum signals.

Three bands are well distinguishable; however, a specific assignment is not possible due to the overlap between the signals of the phenyl ring and the imidazolium ring carbons. At 62 ppm a band, related to the quaternary carbon **1,** is evident, whereas, the band at 24.9 ppm is ascribed to the methyl groups carbons **11** and **9**. $^{13}C$ carbonyl signal **10** of the acetate anion can be visible as a small and broad band at around 200 ppm. It is worth noticing that the main inconvenience to be avoided during the synthesis of **CB-PCP-1** is the formation of a polyimine network, in which the closure of the imidazolium ring by formaldehyde is not fully performed. In this specific case, two $^{13}C$ strong bands between 150 and 200 ppm, related to imine (C=N) $^{13}C$ should be present, as evidenced by Thiel *et al* [83]. **CB-PCP-1** $^{13}C$ CP-MAS solid state NMR spectrum does not show the two bands related to the imine, indicating the successful formation of the imidazolium network. The weak band at 160 ppm can be ascribable to the acetamide moiety present at the chain terminal.[60]

## High resolution Mass spectroscopy

Since sample **1** is obtained in solution after the reaction, we tried to confirm the structure by means of high resolution mass spectroscopy (HR-MS).[61] Unfortunately, the mass weight of the crosslinked polymeric chain of the sample was too high and the electrospray ionization (ESI) failed to transfer the sample from the solution to the gas phase. Therefore we synthesized a partially linear polymer (sample **1s)**, using a lower ratio of aldehydes in respect to the amino monomer (details about this synthesis are reported in the supporting information section), and analysed the reaction solution using HR-MS. From the HR-MS spectra reported in Figure S2**,** it is evident the presence of the dimer in reaction solution having an *m/z* of 809.4156. Furthermore there is a distribution of species ascribable to the dimer bearing the acetamide moiety on the amino groups (Table S2). The formation of the acetamide moiety is induced by the presence of an excess of acetic acid in solution that reacts with the free amino group of the dimer. Trimer and tetramer formed during the reaction are also visible in the HR-MS spectra (Figure S3 and S4), along with their distribution bearing the acetamide moiety on the amino group (Table S3 and S4), proving the growth of the polymeric chain. Polymeric chains longer than tetramer are difficult to be observed because of the high molecular weight, which prevents the de-solvation and the ionization operated by the ESI source.

Further evidences of the closure of the imidazolium are furnished by the $MS^n$ fragmentation obtained for the dimer having *m/z* 809.4156, which presented, together with the purposed chemical structure, in Figure S5 part a-d. The fragmentation shows the loss of the four phenylamino groups, the first two in form of radicals, and the second in form of aniline. The imidazolium ring is not fragmented as it is the strongest part of the molecule due to the system of conjugated bonds. The $MS^n$ fragmentation is also performed on the dimer with one acetamide moiety (m/z of 851.4181 shown in Figure S6), evidencing both the losses of free phenylamino group, and phenylamino group with the acetamide moiety. These two signals exhibit different intensities due to statistical distribution.

## Investigation on CB-PCPs morphological properties

The pore structure of CB-PCPs were investigated by means of $N_2$ adsorption at 77 K. The adsorption isotherms for sample **1** to **6** are reported in Figure 4. All isotherms possessed a characteristic behavior, already observed in case of other polymeric networks.[62, 63] More in detail, samples **1-5** adsorption isotherms show a pronounced knee at low pressures, typical of micropores, then the isotherm profile constantly rise after the micropores filling, without reaching a plateau. A peculiar hysteresis can be observed in all the polymers, where the desorption branch of the isotherm does not close to the adsorption branch at low relative pressures. These phenomena, particularly evident for samples **3**, **4** and **6**, are mostly attributed to a swelling of the polymeric matrix due to the elastic deformations occurring during nitrogen adsorption. [62, 63] Therefore, the porosity of these materials is better described by the desorption branches of the isotherms. In fact, being the swelling of the materials proportional to the gas pressure, a fraction of pores is available for $N_2$ adsorption only at high values of relative pressure.[62, 63] For sample **2**, the desorption branch exhibits a weak step at 0.5 $p/p_0$, typical of materials with slit pores. In the case of samples **1** and **5** the hysteresis is less evident than for the other polymers. The difference in the hysteresis among different CB-PCPs probably comes from the different anions, which can slow down the swelling of the material, operated by $N_2$, during the adsorption step.

The Brunauer-Emmett-Teller (BET) specific surface areas (SSAs) of the different CB-PCPs are reported in Table 1. The microporosity of these materials arises from the inefficient packing of the sterically hindered tecton monomers that induces empty spaces inside the network, therefore high accessibility to the formed imidazolium moieties. A fraction of mesopores is also present, deriving from the fragmentation of the network. Since the adsorption isotherm shows a mixed behavior of mesoporous and microporous materials the value of BET and Langmuir specific surface area (ssa) are both reported in Table 1. In all the materials having the same polymeric backbone, the specific surface area strongly depends on the anions nature. Sample **1**, **2** and **5**, having hydrophilic anions, exhibit the highest SSAs, probably due to the higher capacity of swelling in aqueous solution that leads to a better retaining of the porous structure during the freeze-drying process. On the contrary, samples **3** and **4**, with hydrophobic anions, tend to swell less when suspended in water and after the freeze drying, they show less porosity since the polymeric chain are more entangled. Sample **6,** the CB-PCP bearing carbene, shows a remarkable lower SSA (BET = 77 $m^2/g$) even if its isotherm retains the profile described for the other materials pointing out a micro-/mesoporous character. The decrease of the surface area is probably ascribable to a partial Wanzlick equilibrium occouring during the carbene synthesis, which leads carbenes to couple each other.[64] Furthermore, the different solvent employed for freeze-drying this material (1,4-dioxane) could confer different swelling properties to the polymeric chains.

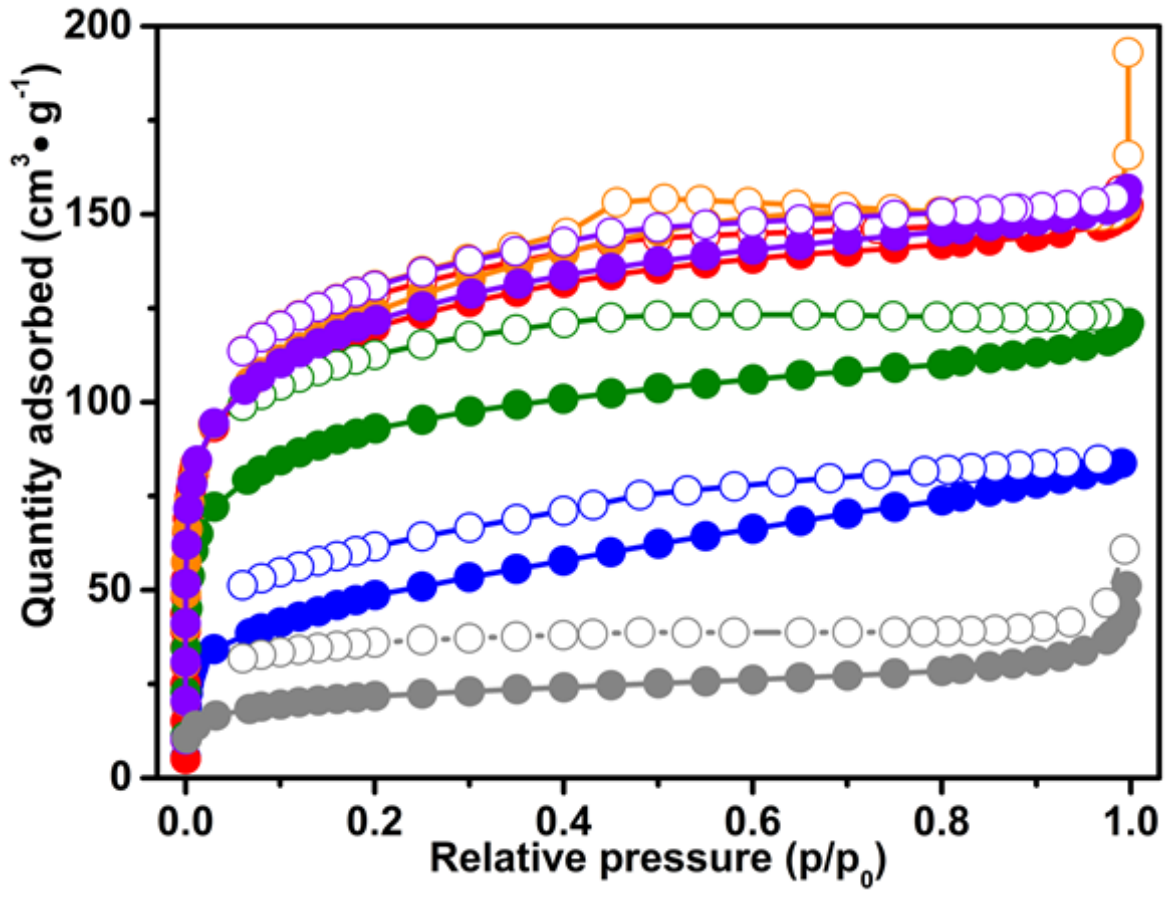

**Figure 4.** $N_2$ adsorption isotherms at 77K for samples: **1** (red curve), **2** (orange curve) **3** (blue curve), **4** (green curve), **5** (violet curve) and **6** (dark grey curve).

**Table 1.** BET and Langmuir specific surface area of sample **1** to **6** and **1a**, **1b**.

| sample name | SSA BET ($m^2/g$) | SSA Langmuir ($m^2/g$) | $V_{total}$ [a] ($cm^3/g$) | $V_{micro}$ [b] ($cm^3/g$) |
|---|---|---|---|---|
| 1 | 419 | 570 | 0.203 | 0.131 |
| 2 | 436 | 595 | 0.187 | 0.109 |
| 3 | 176 | 245 | 0.109 | 0.034 |
| 4 | 325 | 441 | 0.146 | 0.088 |
| 5 | 426 | 578 | 0.187 | 0.112 |
| 6 | 77 | 105 | 0.055 | 0.013 |
| 1a | 396 | 530 | 0.128 | 0.120 |
| 1b | 8.7 | 12.9 | - | - |

[a] Total pore volume obtained from NL-DFT analysis of the adsorption isotherm.
[b] Micropore volume obtained from NL-DFT analysis of the adsorption isotherm.

The pore volumes were derived from the adsorption branches of the isotherms, using the non-local density functional theory (NL-DFT) pore model for carbon with slit pore geometry. The total pore volume is around 0.19 $cm^3/g$ for samples bearing hydrophilic anions (*i.e.* sample **1**, **2** and **5**) while it is lower for samples with inferior surface area. The ratio between the micropore volume and the total pore volume is approximately constant (~60%) for all the samples, except for sample **3** which presents a lower ratio (~30%). This behavior can be ascribed to the steric hindrance of the large, hydrophobic $Tf_2N^-$ anion, which can occlude the micropores. The total pore volume is lower for sample **6** due to the lower SSA, and the sample has a low micropore/total pore volume ratio (~24%), probably due to the increased crosslinking density arising from coupling of carbenes.

In order to investigate the effect of the starting solution concentration, on the textural properties of the resulting materials, samples **1a** and **1b** have been synthesized employing solutions two or four times respectively more concentrated (synthesis details are reported in supporting information). The isotherms of these samples are shown in Figure S7, whereas the values of SSA and pore volume are reported in Table 1. Sample **1a** shows a a BET SSA very similar to the value reported for sample **1** (396 $m^2/g$). However, for this material, almost all the pores volume consists of micropores. This behavior could derive from the polymer gelification that can occur employing a lower reaction solution volume. The decrease of the available reaction volume induces a reduced swelling process of the polymer that, in turn, is balanced by the steric repulsion of the tecton monomers, producing an extensive polymeric network mainly constituted of micropores. In the case of sample **1b,** the further lower reaction volume generates a bulky polymer with an almost null SSA (Figure S7). Pore size distribution can be found in Figure S8 and S9.

Samples morphology was investigated by SEM. Representative SEM images of sample **1** are reported in Figure 5 and in Figure S10 (lower magnification). CB-PCP **1** appears in form of particles of 1-20 µm. The particles are irregular in shape with evident fragmentation clearly visible in Figure 5. The microporous structure of these polymers is instead not visible with SEM. The hierarchical porosity in CB-PCPs allows an easy diffusion of gas molecules inside the polymeric matrix thus reaching the imidazolium active sites. SEM images of the other CB-PCPs are not reported being these materials morphologically very similar to sample **1**.

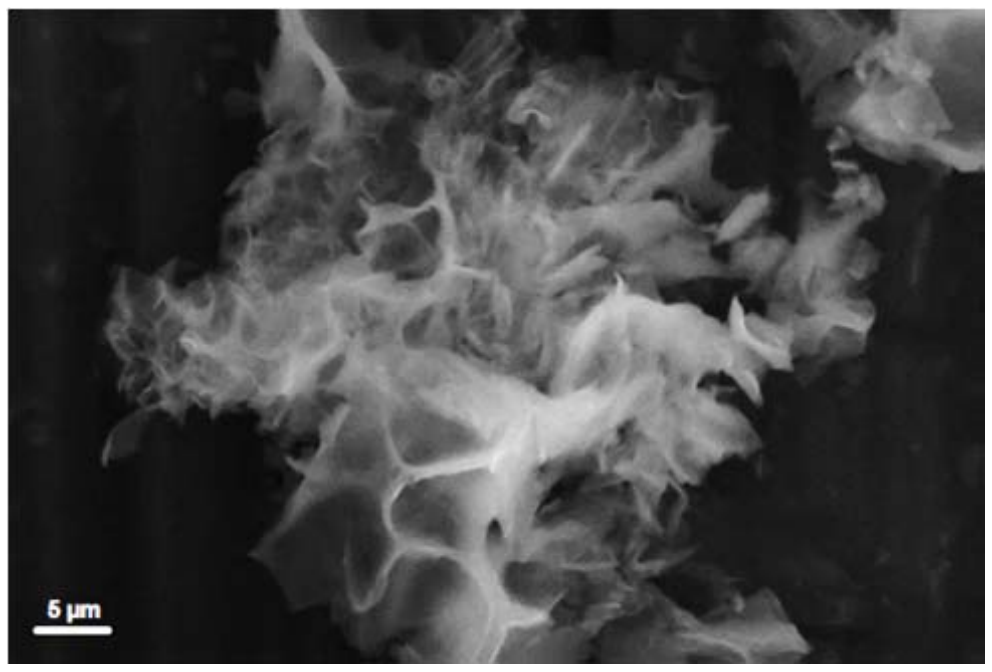

**Figure 5.** SEM picture of sample **1.**

## Thermogravimetric analysis

The thermogravimetric analysis of the synthesized CB-PCPs **1**, to **5** are reported in Figure S11. All the thermogravimetric curves are similar despite the different anions, showing a drastic step of weight loss starting at around 400°C related to the decomposition of the polymeric network. The decomposition step is sharp, probably due to the irregular structure of the network, which broadens the thermal energies range at which the polymer chains break. The materials also exhibit one small weight-loss step starting at 110°C, ascribable to the removal of residual moisture. It is worth noticing that the thermal degradation of the CB-PCPs does not depend on the anion nature. This behavior highlights that the thermal degradation pathway is not the de-alkylation of the imidazolium nitrogen, as observed for dangling imidazolium PILs, but rather the decomposition of the entire polymeric network in one step.

## Carbon dioxide adsorption study on CB-PCPs

The presence of the imidazolium ionic liquid moiety, well known to strongly interact with the $CO_2$ molecule, together with a microporous network, makes CB-PCPs promising materials for $CO_2$ capturing. For this reason, a series of volumetric adsorption measurements up to 1 bar and at different temperatures have been performed on the obtained CB-PCPs. Adsorption and desorption isotherms, recorded at 298 K for samples **1**-**6**, are reported in Figure 6. All samples exhibit the peculiar hysteresis observed also for $N_2$ adsorption at 77K, in which the desorption branch of the isotherm does not close to the adsorption branch even at low relative pressures. The $CO_2$ uptake at 298 K and 1 bar is between 1 and 1.2 mmol/g for all the materials. In particular, sample **1** shows the highest loading with a value of 1.2 mmol/g, whereas sample **3** shows the lowest loading of 0.9 mmol/g. The most pronounced hysteresis, observed in the case of sample **3,** can be ascribable to the steric hindrance of the large, hydrophobic $Tf_2N^-$ anion that limits the access of the carbon dioxide to the imidazolium moiety. Anyhow, all samples show a complete release of carbon dioxide at room temperature. Another general feature, is the fact that all the isotherms do not reach a plateau at 1 bar. This observation encourage further studies at higher pressures extending their interest to pre-combustion $CO_2$ capture.[51]

The $CO_2$ uptake was also evaluated for sample **1** at 273 K and 313 K. The isotherms reported in Figure S12 highlight only a slight decrease of the adsorption capacity at 313 K with a maximum loading of 0.95 mmol/g, on the contrary, at 273 K, the performance of the material drastically improves, almost doubling the maximum loading to a value of 2.05 mmol/g. To the best of our knowledge this value attests the herein described CB-PCPs as the most performant $CO_2$ adsorbent materials nowadays in the field of imidazolium PILs. All the previous studies involving PILs for carbon dioxide adsorption are listed in an extensive review from Zulfigar *et al.*, which encompass PILs with imidazolium, pyridinium or tetraalkylammonium cation and a variety of anions.[48] According to this review the best $CO_2$ adsorption at 1 bar was measured in case of a cross-linked mesoporous imidazolium PILs obtained by silica hard-templating pathway (0.46 mmol/g) (P(SVImTf$_2$N)).[65] In the field of porous imidazolium polymers, it is remarkable the work reported by Zhao *et al.* in which the cross-linking between the polymeric chains is obtained in a template-free synthesis via complexation between anion and cation both present along the main chain ; in this case the reported $CO_2$ adsorption at 1 bar was 0.64 mmol/g (P(CMVImBr1.03-co-AA)).[66] Other two works are also noteworthy reporting silica supported tetraalkylammonium PILs ($SiO_2$-P(VBTMA)($BF_4$)) and linear main-chain anionic PILs, having 1-butyl-3-methylimidazolium as counteraion (PUA-02), both reporting maximum carbon dioxide adsorption of around 0.4 mmol/g at 1 bar.[67, 68] One recent work dealing with cross-linked di-vinylimidazolium PILs reports $CO_2$ adsorption value of 1.02 mmol/g at 273 K and 1 atm (PDMBr).[69] Whereas one study by Talapaneni *et al.* dealing with imidazolium porous polymer obtained with a two-step synthesis reports adsorption of 1.74 mmol/$CO_2$ at 1 bar and 273 K (NP-imidazolium).[7] This material is very similar to our CB-PCPs having equal SSA, but it has two more phenyl rings connecting every imidazolium moiety, slightly decreasing the ratio between imidazolium functional group and aryl chain, as a consequence, slightly decreasing the carbon dioxide loading.

In order to allow the comparison with the data reported in the literature, the isotherms of $CO_2$ uptake on CB-PCPs have been reported in mmol of carbon dioxide adsorbed per gram of each material (see Figure 6). However, to disclose information on the carbon dioxide adsorption from the molecular point of view, it is more relevant to take into consideration the mol% of carbon dioxide adsorbed respect to the imidazolium moiety, as showed in Figure S13. From these isotherms in fact, it is clear that carbon dioxide adsorption depends on the anion of the CB-PCPs. A previous study reports higher $CO_2$ adsorption for PILs with acetate anion respect to other anions.[70] Furthermore some speculations were reported about a possible generation of imidazolium NHC upon heating and outgassing acetate PILs.[71, 72] Whereas in the current case acetate anion is the one with lowest performance towards $CO_2$ adsorption. Usually, inorganic anion such as $PF_6^-$ or $BF_4^-$, perform better than $Tf_2N^-$ and $TfO^-$ towards carbon dioxide adsorption for PILs developed in linear fashion.[70, 73] However, for cross-linked porous PILs, $Tf_2N^-$ anion showed the best performance.[48, 65] Generally speaking, it is difficult to discriminate among the effect of anion nature, SSA and polymer structure for carbon dioxide adsorption, because the process is a complex of chemical interaction, diffusion kinetics inside the polymer network and steric hindrance of the anion. In accordance with the data already reported for porous PILs, the $CO_2$ adsorption by CB-PCPs follows the same trend with anion: $Tf_2N^- > PF_6^- \approx TfO^- > BF_4^- > AcO^-$.

In comparison to other materials for $CO_2$ adsorption, our CB-PCPs show better adsorption properties than: activated carbon, polycarbazole (PCBZ),[74] porous polymer network (PPN-80),[75] hypercrosslinked polymer (HCP-1)[76] and can compete with common porous aromatic framework (PAF-1).[77] Nevertheless CB-PCPs still show less $CO_2$ uptake than top performer materials like: zeolitic imidazolate framework (ZIF-78),[78] zeolitic tetrazolate framework (ZTF-1)[79], zeolite 13X (ZEO13X),[80] and many metal organic framework.[81-83] Table 2 summarize the carbon dioxide adsorption at 1 bar for poly(ionic liquid)s and other kind of materials discussed in the text, in order to have a direct comparison of these sorbents with respect to CB-PCPs.

**Table 2.** Carbon dioxide adsorption performance of various selected materials

| Sorbent | $CO_2$ mmol·g$^{-1}$ | Conditions (P, T) | Ref. |
|---|---|---|---|
| Metal organic framework (Mg-MOF-74) | 8.02 | 1 bar, 298 K | 83 |
| Zeolitic tetrazolate framework (ZTF-1) | 5.59 | 1 bar, 273 K | 79 |
| Metal organic framework (UTSA-16) | 4.90 | 1 bar, 273 K | 82 |
| Zeolite 13X (ZEO13X) | 4.68 | 1 bar, 298 K | 80 |
| Zeolitic imidazolate framework (ZIF-78) | 3.34 | 1 bar, 273 K | 78 |
| Click-based porous cationic polymer (CB-PCP-1) | 2.05 | 1 bar, 273 K | this work |
| Porous aromatic framework (PAF-1) | 2.05 | 1 bar, 273 K | 77 |
| Poly(ionic liquid) NP-imidazolium | 1.74 | 1 bar, 273 K | 7 |
| Hypercrosslinked polymer (HCP-1) | 1.70 | 1 bar, 298 K | 76 |
| Porous polymer network (PPN-80) | 1.62 | 1 bar, 295 K | 75 |
| Polycarbazole (PCBZ) | 1.13 | 1 bar, 273 K | 74 |
| Poly(ionic liquid) (PDMBr) | 1.02 | 1 bar, 273 K | 69 |
| Poly(ionic liquid) P(CMVImBr1.03-co-AA) | 0.64 | 1 bar, 273 K | 66 |
| Poly(ionic liquid) P(SVImTf2N) | 0.46 | 1 bar, 273 K | 65 |
| Silica-poly(ionic liquid) ($SiO_2$-P(VBTMA)($BF_4$)) | 0.40 | 1 bar, 303 K | 67 |
| Poly(ionic liquid) (PUA-02) | 0.40 | 1 bar, 298 K | 68 |

Generally speaking, good performances towards $CO_2$ adsorption in porous polymer are obtained from a combination of high surface area, microporosity, and high concentration of imidazolium active sites along the polymeric backbone. Furthermore the anion is an integrating part of the adsorption properties and its choice is not straightforward, but depends on the chemical and morphological structure of the polymeric backbone. Remarkable is the weight ratio between the imidazolium moiety and other parts in the polymer chemical structure, which is very high in CB-PCPs. Furthermore the direct conjugation of the imidazolium ring with two phenyl groups can change the distribution of its positive charge, then affecting the way imidazolium interacts with $CO_2$. It is worth noticing that, even though the CB-PCPs have about one tenth of the SSA of PAF-1 (5640 $m^2$/g),[15] they show very similar $CO_2$ capture capacities, testifying the relevance of the introduction of functional groups highly dispersed and accessible. In order to evaluate the effect of the porosity of CB-PCPs on carbon dioxide loading, the $CO_2$ adsorption at 298 k and 1 bar has been performed also on morphologically different samples **1a** and **1b**. The adsorption isotherms are reported in Figure S14. Sample **1** and **1a** exhibit exactly the same loading of 1.2 mmol/g at 1 bar; furthermore, sample **1a** shows a less pronounced hysteresis, probably due to the higher uniformity of the material in terms of microporosity.

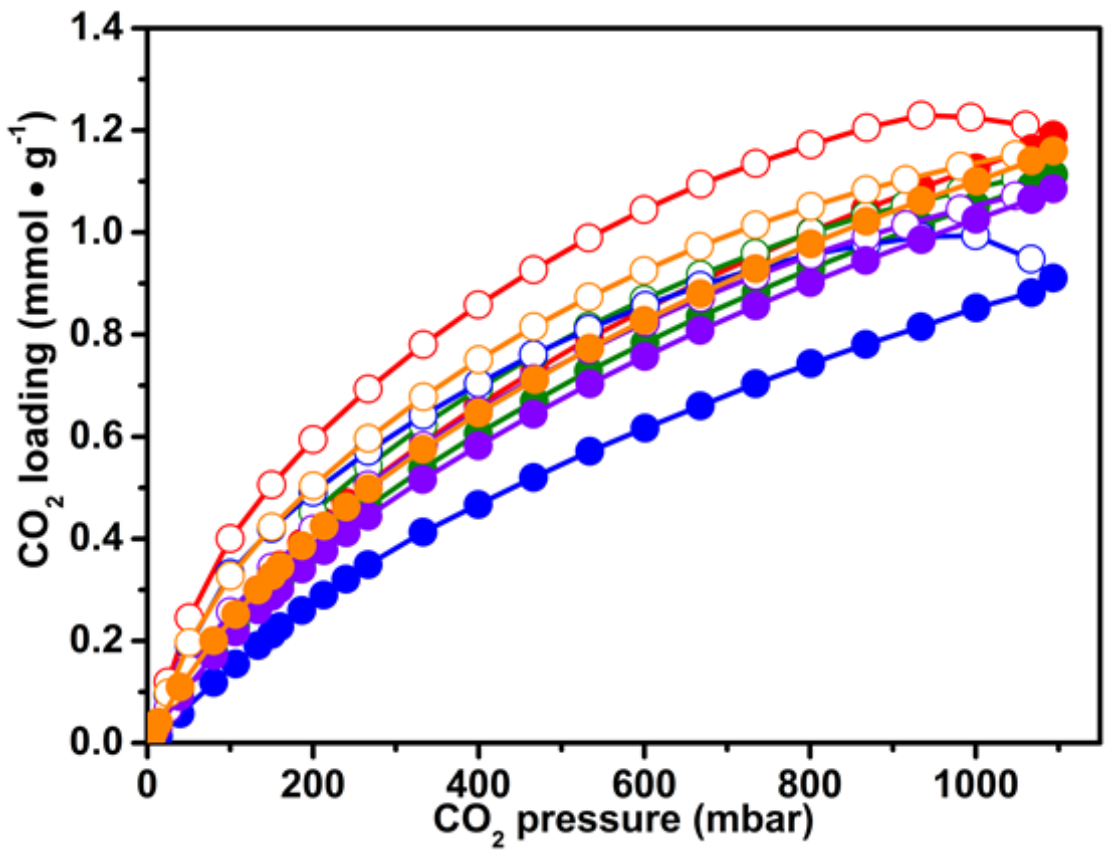

**Figure 6.** Carbon dioxide adsorption isotherm at 298 K for samples: **1** (red curve), **2** (blue curve), **3** (green curve), **4** (orange curve), **5** (violet curve). Spheres describes the adsorption branch, while circle describe the desorption branch.

Sample **1b,** that has almost no surface area (see Figure S7 and Table 1), shows however good adsorption towards carbon dioxide, with a maximum loading of 1 mmol/g. In fact, carbon dioxide can acts as plasticizer for PILs and it can, in part, penetrate the bulk structure of non-porous polymers in a long diffusion time.[84, 85] For this reason, it is clear that the nature of anions is more relevant than SSA in determining the $CO_2$ adsorption capacities of these materials.

## Carbon dioxide capture in presence of the carbene moiety

It is well-known that by means of strong organic bases, it is possible to introduce N-heterocyclic carbene on the C2 position of imidazolium ring.[86-88] This carbene, in turn, can react with carbon dioxide forming a new C-C bond and a carboxilate directly linked to imidazolium ring. This imidazolium carboxilate is thermally unstable and it can decompose below 100 °C releasing carbon dioxide and restoring the carbene, which can restart the adsorption/desorption cycle.[89-91] The overall process is reported in Scheme 2. The chemical looping is considered as a new frontier in carbon dioxide adsorption.[92] This mechanism was already studied for different ionic liquids and poly(ionic liquid)s, as a way for storing carbon dioxide and also protecting the carbene from the decomposition in the presence of moisture.[7, 66, 89, 90, 93-97]

In our study we direct efforts to unravel the differences from the physico-chemical point of view between carbon dioxide capture operated by the as-synthesized CB-PCPs and CB-PCPs bearing NHC. A direct comparison between the $CO_2$ adsorption capacity at 298 K of CB-PCPs bearing NHC (sample **6**) and its precursor CB-PCP (sample **1**) is reported in Figure 7. The maximum $CO_2$ loading of sample **6** is 1.1 mmol/g, a value perfectly comparable with the uptake reported for sample **1**. Conversely, from a molecular point of view, data reported in mol% of $CO_2$ loading respect to the imidazolium or NHC moieties clearly show that the maximum loading drastically decreases after the introduction of the carbene (from 36 mol% for sample **1** to 26 mol% for sample **6**). This behavior is ascribable to the lower number of available active sites. In fact, when NHC carbene are formed they are so reactive that tent to couple together forming a dimer. This phenomenon is also responsible for the evident decrease of the SSA of sample **6** (see Table 1). The formed dimers, without free electronic doublet, are not chemically able to bond carbon dioxide, thus decreasing sorption capacity respect to imidazolium moiety. However, the lightness of sample **6**, arising from a lower molecular weight of the monomeric unit, compensates the loss of active sites, giving the same performance in terms of carbon dioxide loading per mass unit of adsorbent.

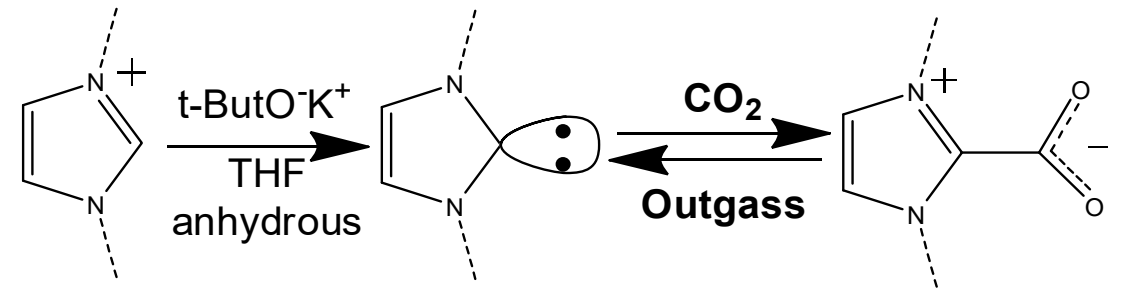

**Scheme 2.** Synthetic step for the introduction of carbene in the imidazolium ring (left side) and reversible formation of the NHC-$CO_2$ adduct (right side).

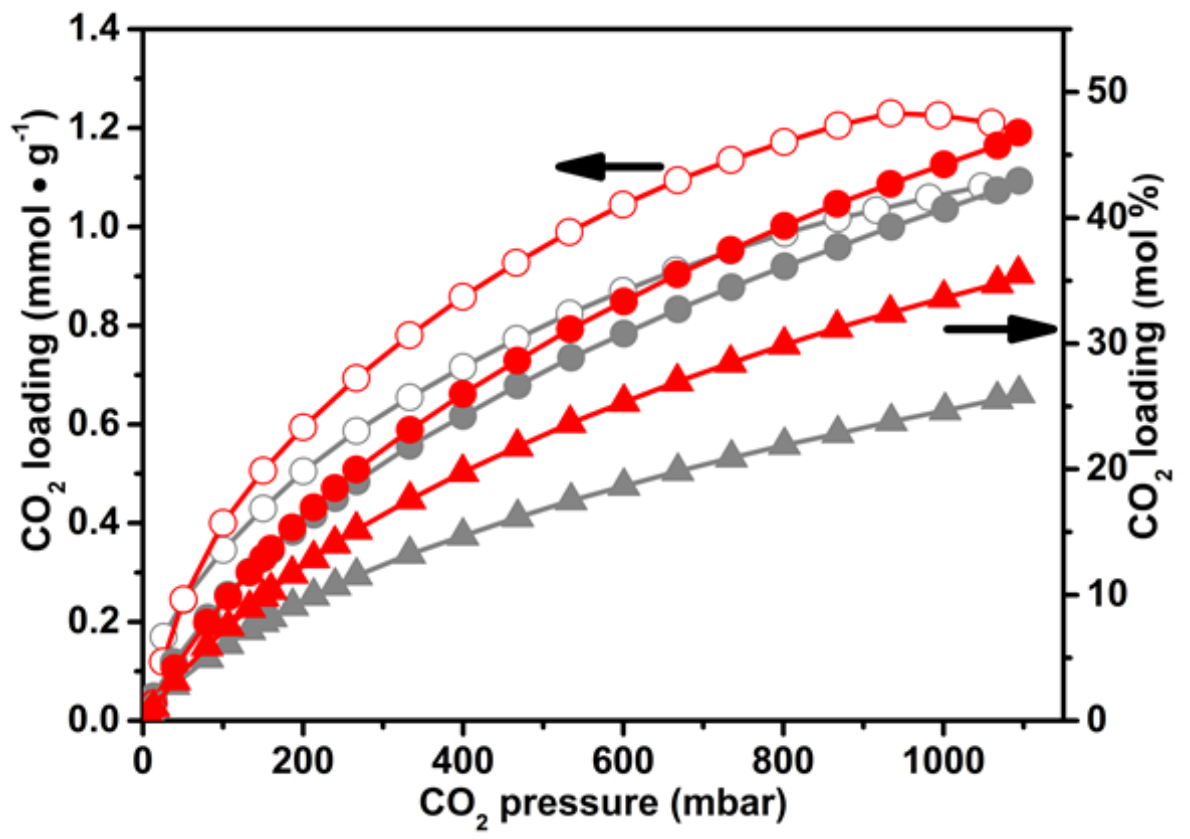

**Figure 7.** Carbon dioxide adsorption isotherm at 298 K for sample **1** and **6**. Spheres and circle describes respectively the adsorption and branch referring to the scale on the left (mmol/g), while triangles of adsorption branch refers to to the scale on the right (mol% of $CO_2$ respect to the imidazolium or NHC moieties).

*In-situ* FTIR spectroscopy was used to follow the reaction between the polymer bearing NHC and carbon dioxide. The effect of carbon dioxide contact on pre-activated sample **6**, is illustrated in Figure 8. The IR spectrum of sample **6** after activation (black curve) is characterized by the spectral features of the polymeric framework. Upon 1 h contact with 200 mbar of $CO_2$, two new bands appear in the spectrum at 1665 $cm^{-1}$ and at 1295 $cm^{-1}$ (dark grey curve). ascribable to the $\nu_{asym}(OCO^-)$ and the $\nu_{sym}(OCO^-)$ of the formed imidazolium carboxylate.[66, 89, 98, 99] The formation of a carboxilate testify the activation of carbon dioxide, as illustrated by the deep change in its molecular orbitals.[100] It is worth noticing that, after the $CO_2$ contact, the evacuation at room temperature (light grey curve) almost restore the spectrum of the material after activation. The same experiment was repeated for sample **1** without any evidence of carboxylate formation upon exposure of the CB-PCP at 200 mbar of carbon dioxide.

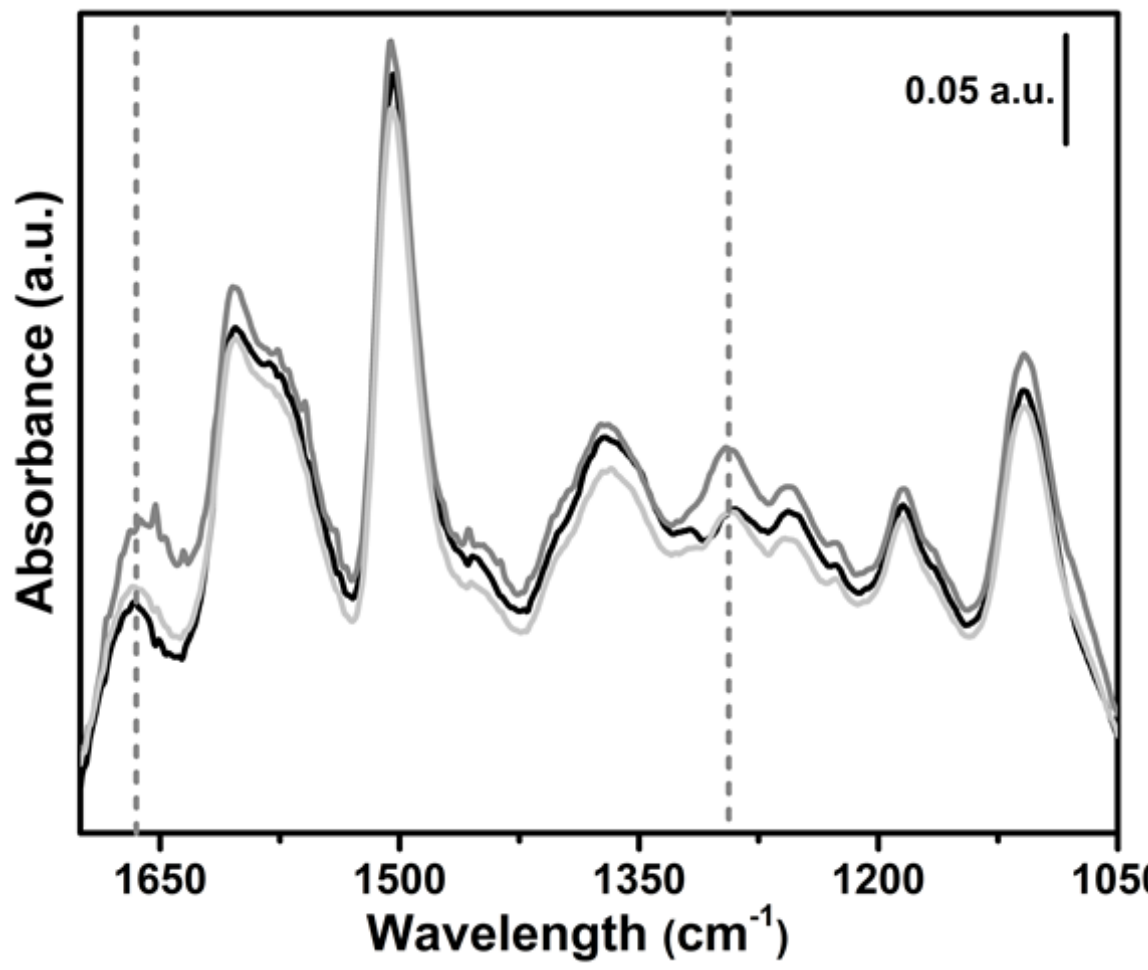

**Figure 8.** *In-situ* FTIR spectra upon dosage of 200 mbar carbon dioxide on sample **6**. Black curve: activated sample. Dark grey curve: 1 h contact with 200 mbar of $CO_2$. Light grey curve: $CO_2$ evacuation.

Micro-calorimetric gas adsorption experiment were performed to further investigate the interaction energy between carbon dioxide, ionic CB-PCP, and their carbene counterpart. The differential molar adsorption heats of samples **1** and **6** are reported as a function of carbon dioxide coverage in Figure **9**, whereas their corresponding quantitative isotherms are shown in Figure S15. In the case of sample **1,** the differential heat of adsorption at zero coverage is 35 KJ/mol, then it decreases to 10 KJ/mol for higher $CO_2$ coverages, *i.e.* significantly under the value of $CO_2$ molar liquefaction enthalpy. This low isosteric heat could be explained considering two overlaying contributions: the gas adsorption and the structure rearrangement as a consequence of the polymer swelling process.[66] Conversely, for sample **6** the differential heat of adsorption is substantially higher starting from 58 KJ/mol at low coverage and decreasing to about 35-40 KJ/mol at high coverage. The higher differential heat of interaction of sample **6** is perfectly in line with the adduct formation between NHC and $CO_2$. In fact, the creation of the C-C covalent bond largely compensate $CO_2$ deformation. The differential heat of adsorption is almost the same for the primary and secondary adsorption. Whereas in Figure S16, which reports the integrating heat of adsorption as function of $CO_2$ pressure, it is possible to see that the secondary adsorption exactly matches the primary for sample **1**, while in the case of sample **6**, a small differences in the integrating heat is present. Micro-calorimetry underline the almost complete reversibility of carbon dioxide adsorption, both in the case of ionic polymer (Sample **1**) and in the case of NHC polymer (Sample **6**).

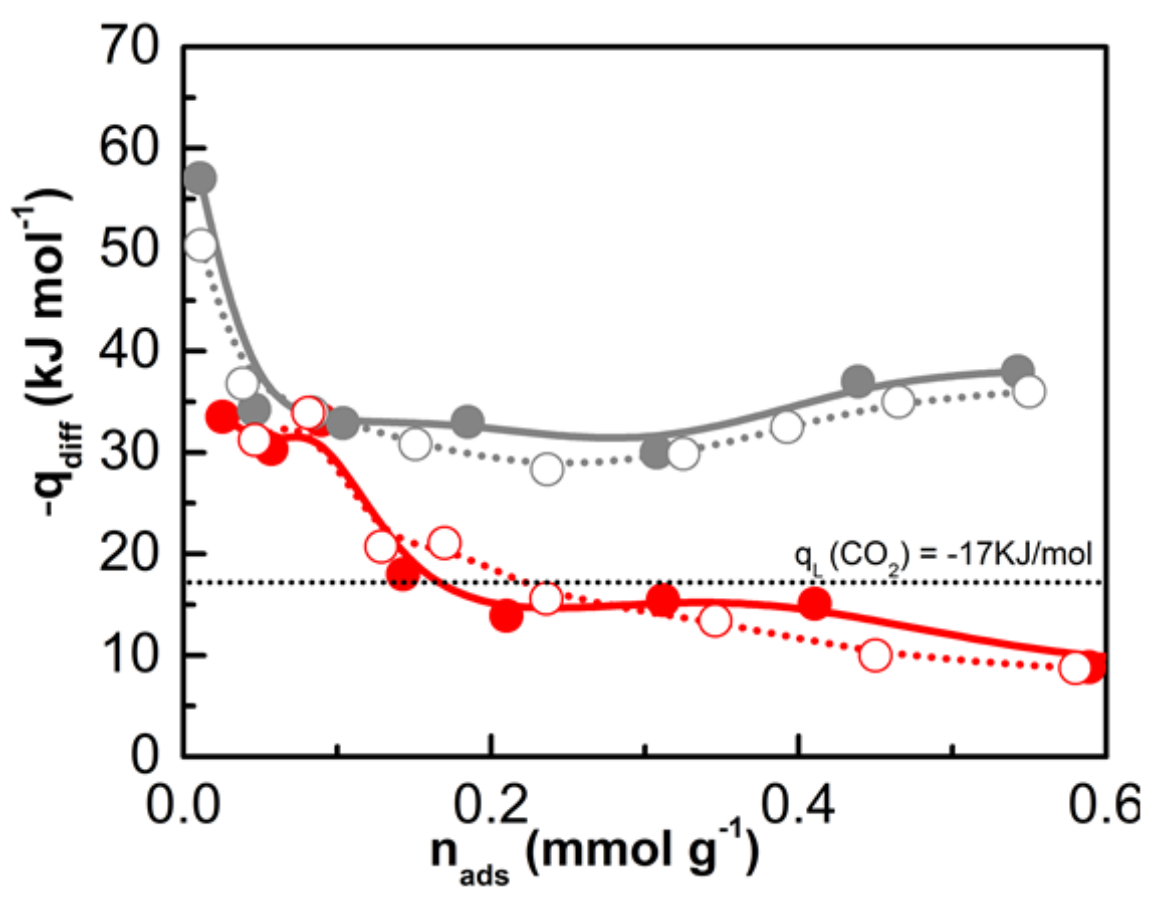

**Figure 9.** Differential molar adsorption heatsas a function of the coverage relative to the adsorption at 298 K of $CO_2$ on samples **1** (red curves) and **6** (grey curves). Spheres refer to the primary adsorption adsorption and circles to the secondary ones. The dotted horizontal line represents the standard molar enthalpy of liquefaction of $CO_2$ at 298K.

## Conclusions

We described a facile and straightforward synthetic way to obtain click reaction-based micro-/mesoporous cationic polymers. A set of materials with different anions was successfully synthesized and characterized in order to confirm their structure and study the porosity. It was demonstrated that CB-PCPs exhibit an excellent behavior towards carbon dioxide adsorption, either at 298 K and 1 bar, where the loading is above 1 mmol/g for all the samples, or even more at 273K and 1 bar where the loading for CB-PCP **1** is 2 mmol/g, achieving the highest value for porous ionic polymer. CB-PCP **1** was also modified in order to introduce NHC on the imidazolium ring. The performances towards carbon dioxide adsorption for NHC CB-PCP are of the same level of that ones for ionic CB-PCPs, even though they show a different adsorption mechanism. The combination of *in-situ* FTIR spectroscopy and of adsorption micro-calorimetry reveals two adsorption processes of completely different nature: on one hand, a plain physisorption process in the ionic CB-PCPs and, on the other hand, the chemisorption of carbon dioxide of NHC polymer.

## Acknowledgements

Thanks are due to Prof. Leonardo Marchese and Geo Paul (Università del Piemonte Orientale "A. Avogadro") for SS-NMR measurements and the useful discussions.

## Funding Sources

Italian Ministry MIUR-(project PRIN 2010-2011 n:2010A2FSS9). ERC (European Research Council) Starting Grant. Project number 639720 – NAPOLI

# Supporting Information

## Click-Based Porous Cationic Polymer for Enhanced Carbon Dioxide Capture

Alessandro Dani*, Valentina Crocellà, Claudio Magistris, Valentina Santoro, Jiayin Yuan, Silvia Bordiga*.

**Table of Contents**

## 1 Materials

All solid and liquid starting materials were purchased from Sigma-Aldrich and used as received.

### *1.1 Synthesis of CB-PCP 1*

The general procedure for the synthesis of CB-PCP, showed in Scheme 1 is the following: in the first vial tetrakis(4-aminophenyl)methane (0.228 g, 0.600 mmol) was dissolved in 16 mL of a solution of water:glacial acetic acid 50:50. In the second vial formaldehyde 37% solution in water (0.394 mL, 5.28 mmol) and methyl glyoxal 40% solution in water (0.812 mL, 5.28 mmol) were mixed with 2 mL of water and 1 mL of glacial acetic acid. The solutions of the two vials were mixed together and the solution immediately turned form transparent to dark yellow, the reaction ran for 12 hours at 80°C and the solution became dark brown. The polymer was purified by dialysis against MilliQ® water using the 3.5 kDa tubing, afterwards it was freeze-dried. The obtained CB-PCP material, named **1**, was in form of brown fine powder.

### *1.2 Anion exchange on CB-PCP 1*

The general procedure for the anion exchange, represented in Scheme 1 is the following: the polymer solution, obtained as described before and prior to the dialysis, was mixed with 20 mL of a solution of sodium tetrafluoroborate ($NaBF_4$, 0.329 g, 3.00 mmol). The polymer immediately precipitate out form the solution, because the change of polarity induced from the anion exchange with more hydrophobic counterion, the precipitated polymer was gently agitated for 1 hour and then washed with water (3x20mL). The exchanged CB-PCP was then freeze-dried and the product, named **2**, is obtained in form of dark brown powder. The same procedure was followed for CB-PCP with other counterion using respectively a solution of: bis(trifluoromethane)sulfonimide lithium salt ($Tf_2NLi$, 0.758 g, 2.64 mmol); potassium hexaflorophosphate ($KPF_6$, 0.486 g, 2.64 mmol); and sodium trifluoromethanesulfonate (NaTfO, 0.454 g, 2.64 mmol). In all the cases, due to the hydrophobicity of the exchanged counterion, the polymer precipitate out from the solution. Samples were respectively named **3** ($Tf_2N$), **4** ($PF_6^-$), and **5** ($TfO^-$).

### *1.3 CB-PCP Carbene synthesis*

CB-PCP **1** was used as starting material to introduce NHC carbene in the CB-PCP. The procedure, represented in Scheme 1c is the following: CB-PCP **1** (0.300 g) was hermetically sealed in a vial and flushed with $N_2$ in order to desorb the adsorbed moisture. In another hermetically sealed vial potassium tert-butoxide (0.250 g) was dissolved in 6 mL of anhydrous THF in $N_2$ atmosphere. The solution of potassium tert-butoxide was transferred inside the vial with CB-PCP **1** and stirred for 72 hours at room temperature. The formed CB-PCP NHC carbene, was washed with anhydrous THF (4x10mL), anhydrous methanol (4x10mL) and anhydrous dioxane (4x10mL). Every washing was performed in $N_2$ protected atmosphere in order to avoid the presence of moisture that can react with the NHC carbene deactivating it. The sample was then freeze-dried and the CB-PCP NHC carbene product, in form of bright brown fine powder was stored in glove box under $N_2$ atmosphere. The so obtained material was named **6**.

In order to investigate the effect of the concentration of the starting reagent on the final CB-PCP two samples were synthesized in more concentrated solutions.

### *1.4 Synthesis of CB-PCP 1a*

The general procedure for the synthesis of CB-PCP **1a**, is similar to the one of sample **1**, except for the volume of the solution. In the first vial tetrakis(4-aminophenyl)methane (0.228 g, 0.600 mmol) was dissolved in 8 mL of a solution of water:glacial acetic acid 50:50. In the second vial formaldehyde 37% solution in water (0.394 mL, 5.28 mmol) and methyl glyoxal 40% solution in water (0.812 mL, 5.28 mmol) were mixed with 1 mL of water and 0.5 mL of glacial acetic acid. In this case, a gelation of the entire solution instantly occured forming a dark violet gel. Even though the reaction was immediate, the reaction ran for 12 hours at 80°C in a hermetically sealed vial, in order to obtain a better cross-linking of the network, since gelation process reduced the molecular motions. The gel was then washed with a water:glacial acetic acid 50:50 solution (3x20 mL), then pure water (3X20 mL), afterwards it was freeze-dried. The obtained CB-PCP material was in form of brown fine powder.

### *1.5 Synthesis of CB-PCP 1b*

The general procedure for the synthesis of CB-PCP **1b**, is similar to the one of **1**, except for the volume of the solution. In the first vial tetrakis(4-aminophenyl)methane (0.228 g, 0.600 mmol) was dissolved in 4 mL of a solution of water:glacial acetic acid 50:50. In the second vial formaldehyde 37% solution in water (0.394 mL, 5.28 mmol) and methyl glyoxal 40% solution in water (0.812 mL, 5.28 mmol) were mixed with 0.5 mL of water and 0.25 mL of glacial acetic acid. In this case, a gelation of the entire solution instantly occured forming a dark violet gel. Even though the reaction was immediate, the reaction ran for 12 hours at 80°C in a hermetically sealed vial, in order to obtain a better cross-linking of the network, since gelation process reduced the molecular motions. The gel was then washed with a water:glacial acetic acid 50:50 solution (3x20 mL), then pure water (3X20 mL), afterwards it was freeze-dried. The obtained CB-PCP material was in form of brown fine powder.

### 1.6 *Synthesis of CB-PCP 1s*

The procedure for the synthesis of CB-PCP **1s**, the linear polymer used in the HR-MS, is the following: in the first vial tetrakis(4-aminophenyl)methane (0.0114 g, 0.030 mmol) was dissolved in 0.8 mL of a solution of water:glacial acetic acid 50:50. In the second vial formaldehyde 37% solution in water (0.00462 mL, 0.030 mmol) and methyl glyoxal 40% solution in water (0.00223 mL, 0.030 mmol) were mixed with 0.2 mL of water and 0.1 mL of glacial acetic acid. The solutions of the two vials were mixed together and the solution immediately turned form transparent to dark yellow, the reaction ran for 12 hours at 80°C and the solution became dark brown. The solution so obtained was used for the flow injection analysis in the LTQ-Orbitrap® mass spectrometer.

### 1.7 *Modfied Debus-Radziszewski imidazolium synthesis mechanism*

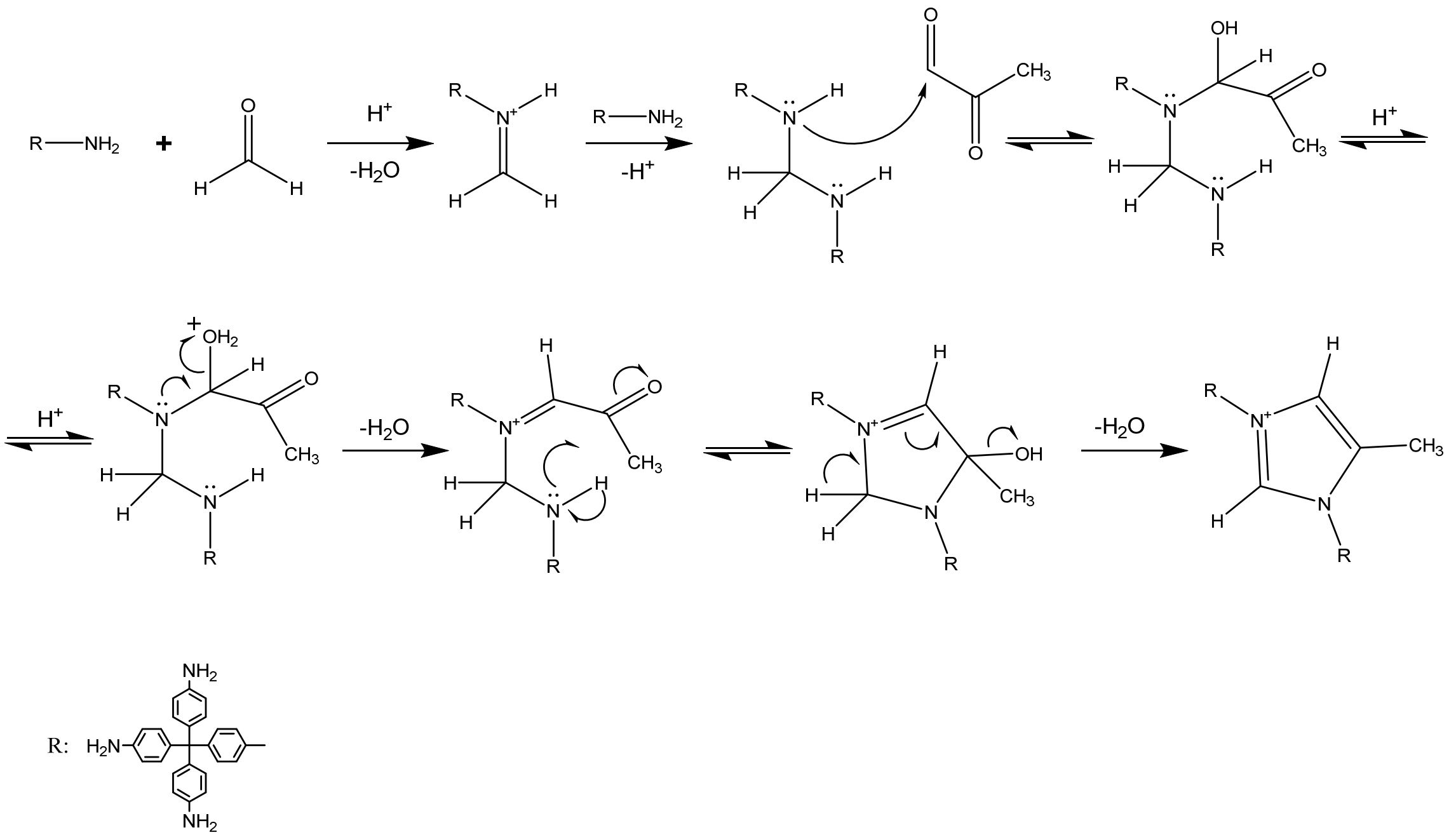

**Figure S1.** Proposed mechanism for imidazolium ring formation.

## 2 Techniques

### 2.1 *Thermogravimetric analisys*

TGA measurements were performed under $N_2$ flow in the range of 30–800˚C (ramp 2˚C $min^{-1}$) using an alumina pan, by means of a TA Q600 analyzer.

### 2.2 *$N_2$ and $CO_2$ adsorption measurements*

$N_2$ adsorption measurements at 77 K and $CO_2$ adsorption at different temperature were performed using Micromeritics ASAP2020. The samples were outgassed for 24 h at 353 K before all measurements. The surface area of the polymers was calculated by means of the BET algorithm and the Langmuir approximation in the standard $p/p_0$ range.

### 2.3 SEM analisys

The morphological properties of the polymers were studied using a SEM Zeiss EVO-50 XV, using a LaB6 source operating at 30 kV current in a high vacuum.

### 2.4 NMR

Solid state NMR (SS-NMR) spectra were acquired on a Bruker Avance III 500 spectrometer and a wide bore 11.7 Tesla magnet with operational frequencies for $^{1}H$ and $^{13}C$ of 500.13 and 125.77 MHz, respectively. A 4 mm triple resonance probe with MAS was employed in all the experiments. The samples were packed on a Zirconia rotor and spun at a MAS rate of 15 kHz. The relaxation delay, d1, between accumulations was between 8.5 s for $^{1}H$ MAS and $^{13}C$ CP-MAS NMR. For the $^{13}C\{^{1}H\}$ CP-MAS experiments, the radio frequency fields $\upsilon_{rfH}$ of 55 and 28 kHz were used for initial excitation and decoupling, respectively. During the CP period the $^{1}H$ RF field $\upsilon_{rfH}$ was ramped using 100 increments, whereas the $^{13}C$ RF field $\upsilon_{rfC}$ was maintained at a constant level. During the acquisition, the protons are decoupled from the carbons by using a TPPM (two pulse phase modulation) decoupling scheme. A moderate ramped RF field $\upsilon_{rfH}$ of 62 kHz was used for spin locking, while the carbon RF field $\upsilon_{rfC}$ was matched to obtain optimal signal and the CP contact time of 2 ms was used. All chemical shifts are reported using δ scale and are externally referenced to TMS at 0 ppm.

### 2.5 High resolution mass spectrometry

Sample CB-PCP **1s** was analyzed by means of LTQ-Orbitrap mass analyzer (Thermo Scientific, Rodano, Italy) equipped with an electrospray ionization source (ESI). The flow injection analysis was performed using a 1000 mg/L solution of Sample **1s** in methanol and a flux of 20µL/min. The analysis was performed in positive ion mode and the parameters of the ESI source were: 16V for capillary voltage and 55V for tube lens. The capillary, magnetic lenses, and collimating multi pole voltages were optimized, and the collision energy (CE) was generally chosen to maintain about 10% of the precursor ion. Mass accuracy was ± 5 ppm. All mass spectra were obtained with resolution of 30000 (500 m/z FWHM). Collision induced dissociation energy (CID) and precursor ion for $MS^{n}$ analyses are written in the top-left part of each mass spectra.

### 2.6 FTIR spectroscopy

ATR-IR spectra were recorded on a Bruker Vertex 70 spectrophotometer equipped with a MCT detector. Each spectrum was recorded at a resolution of 2 $cm^{-1}$ and 32 scans in the range of 4000–600 $cm^{-1}$.

*In-situ* FT-IR spectroscopy analysis was performed on a Bruker Vertex 70 spectrophotometer equipped with a MCT detector. Each spectrum was recorded at a resolution of 2 $cm^{-1}$ and 32 scans in the range of 4000–600 $cm^{-1}$. Sample **6** were manipulated in glove box under $N_2$ atmosphere in order to avoid contact with moisture which can deactivate the carbene. The samples were dispersed in anhydrous methanol, deposited on an IR transparent silicon wafer and then the solvent was evaporated. The silicon wafer with the polymer deposition was then transferred into a special IR cell equipped with two KBr windows. This cell allows the manipulation of the sample in controlled atmosphere while collecting the FT-IR spectra. The cell was linked to a treatment line in order to outgas the sample until for 4 hours at room temperature until $1\cdot10^{-4}$ mbar of residual pressure. After the activation the sample was exposed to growing pressure of carbon dioxide from 0 to 200 mbar and leave in contact with $CO_2$ for 1 hour. At the end of the experiment the sample was outgassed until $1\cdot10^{-4}$ mbar of residual pressure. FT-IR spectra were recorded at each stages.

### 2.7 Adsorption micro-calorimetry

$CO_2$ heats of adsorption were measured, at 298 K, by means of a heatflow microcalorimeter (Calvet C80, Setaram, France) connected to a grease-free high-vacuum gas-volumetric glass apparatus (residual p ≈ $10^{-8}$ Torr) equipped with a Ceramicell 0-100 Torr gauge and a Ceramicell 0-1000 Torr gauge (by Varian), following a well established stepwise procedure. This procedure allows to determine, during the same experiment, both integral heats evolved ($-Q_{int}$) and adsorbed amounts ($n_a$) for very small increments of the adsorptive pressure. Before each measurement the samples, inserted in the calorimetric cell, were outgassed for one night at 353K until $1\cdot10^{-4}$ mbar. The cell was then inserted inside the calorimeter and equilibrated for one night at 298 K.

Adsorbed amounts have been plotted vs. pressure in the form of volumetric (quantitative). The adsorption heats observed for each small dose of gas admitted over the sample ($q_{diff}$) have been finally reported as a function of coverage, in order to obtain the (differential) enthalpy changes associated with the proceeding adsorption process. The differential-heat plots presented here were obtained by taking the middle point of the partial molar heats ($\Delta Q_{int}/\Delta n_a$, kJ/mol) vs na histogram relative to the individual adsorptive doses.

In all quantitative/calorimetric experiments, after the first adsorption run carried out on the bare activated samples (the primary isotherm), samples were outgassed overnight at the adsorption temperature (298 K), and then a second adsorption run was performed (the secondary isotherm), in order to check whether secondary and primary adsorption runs coincided, or an irreversible adsorbed fraction was present.

## 3 Results

### *3.1 Elemental analysis*

**Table S1.** Percentage of C, H, and N of sample **1**, obtained from two repetitions of elemental analysis. Calculated values are reported as reference.

| **entry** | **% C** | **% H** | **% N** |
|---|---|---|---|
| calculated | 74.98 | 5.30 | 9.20 |
| sample **1** | 74.78 | 5.55 | 9.11 |
| sample **1** (repetition) | 74.80 | 5.56 | 9.12 |

### *3.2 Dimer identification from the HR mass spectreometry*

In the mass spectrum of Figure S2 the signals related to the dimer and to the dimer with various amides on the free amine are present. The amide formed through the reaction between the amine and the acetic acid present in the solvent. The amides moieties can form in any free amine and their distribution is statistically, based depending on the concentration of acetic acid in respect to the free amine. Table S2 reports the exact m/z ratio of the dimer and of the dimer with growing number of amides. All the molecules are evident in the mass spectra reported in Figure S2, except for the dimer with all the amino groups in form of amide, because this molecule is not statistically favoured at the used concentration of acetic acid in solution.

**Table S2.** Exact m/z ratio of the dimer, and of the dimer with growing number of amides.

| number of acetamide | m/z |
|---|---|
| 0 | 809.4075 |
| 1 | 851.4181 |
| 2 | 893.4286 |
| 3 | 935.4392 |
| 4 | 977.4498 |
| 5 | 1019.4603 |
| 6 | 1061.4709 |

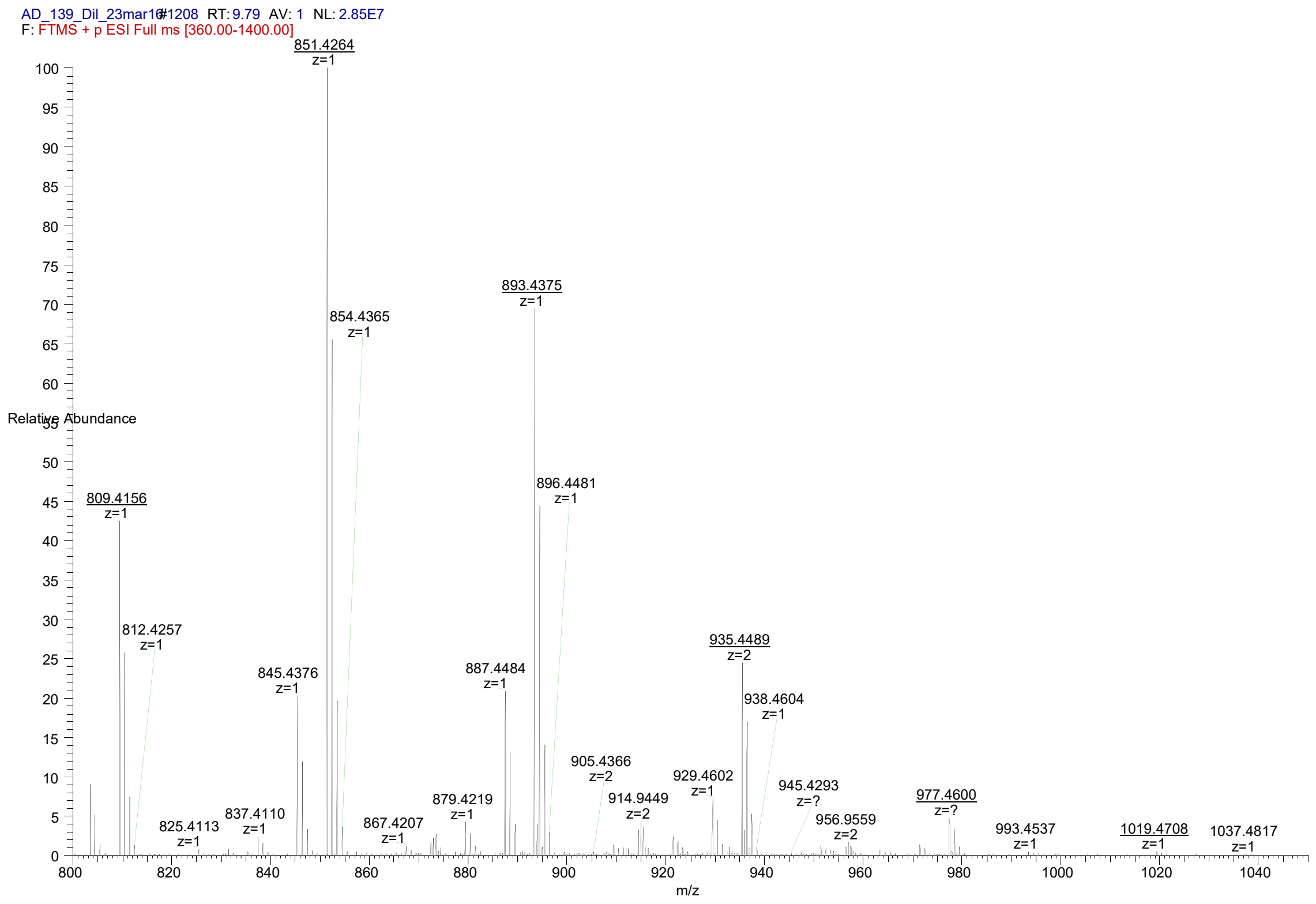

**Figure S2.** HR-Mass spectrum of sample **1s** in the 800-1050 m/z range. The underlined m/z refer to the tetramer in the various amide shapes.

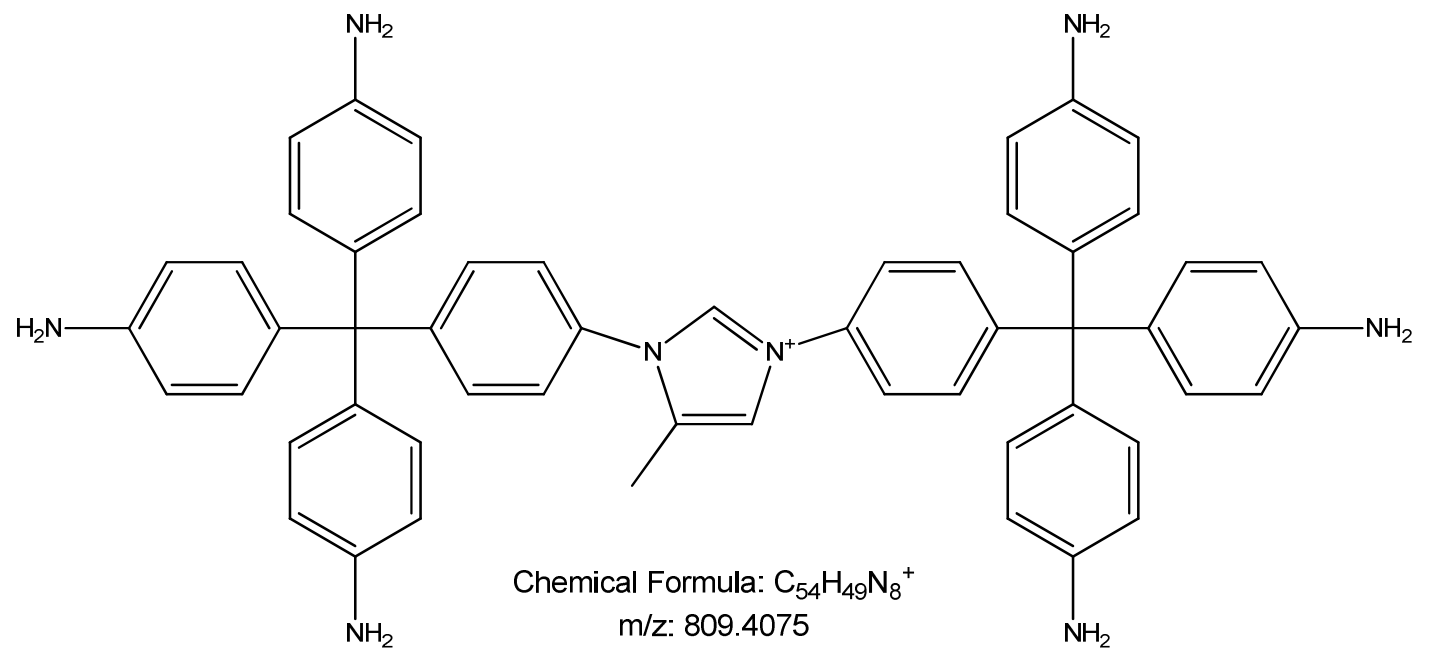

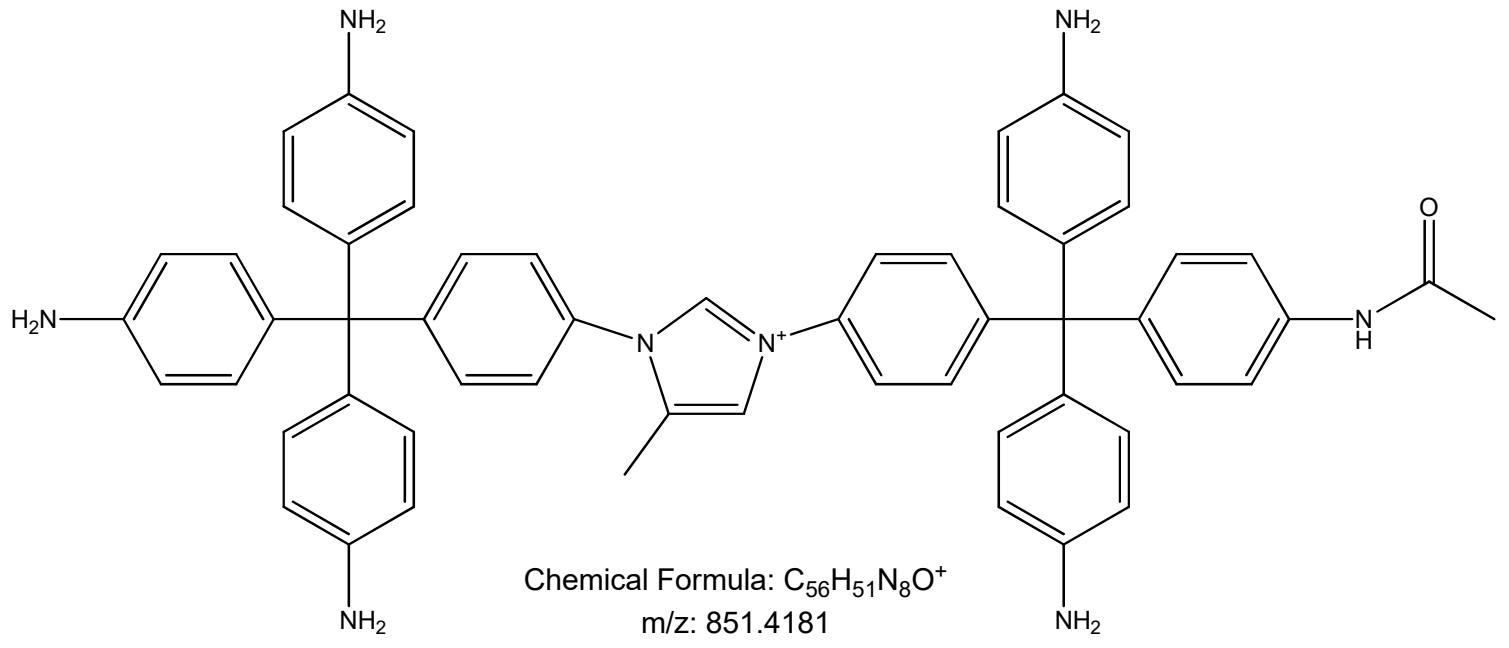

### *3.3 Trimer identification from the HR mass spectrometry*

In the mass spectrum of Figure S3 the signals related to the trimer and to the trimer with various amides on the free amine are present. Table S3 reports the exact m/z ratio of the trimer and of the trimer with growing number of amides moiety. All the molecules are evident in the mass spectra reported in Figure S3, except for the trimer with seven and eight amino groups in form of amide, because these molecules are not statistically favoured at the used concentration of acetic acid in solution.

**Table S3.** Exact m/z ratio of the trimer, and of the trimer with growing number of amides.

| number of acetamide | m/z |
|---|---|
| 0 | 619.3074 |
| 1 | 640.3127 |
| 2 | 661.3180 |
| 3 | 682.3233 |
| 4 | 703.3286 |
| 5 | 724.8355 |
| 6 | 745.8408 |
| 7 | 766.8461 |
| 8 | 787.8514 |

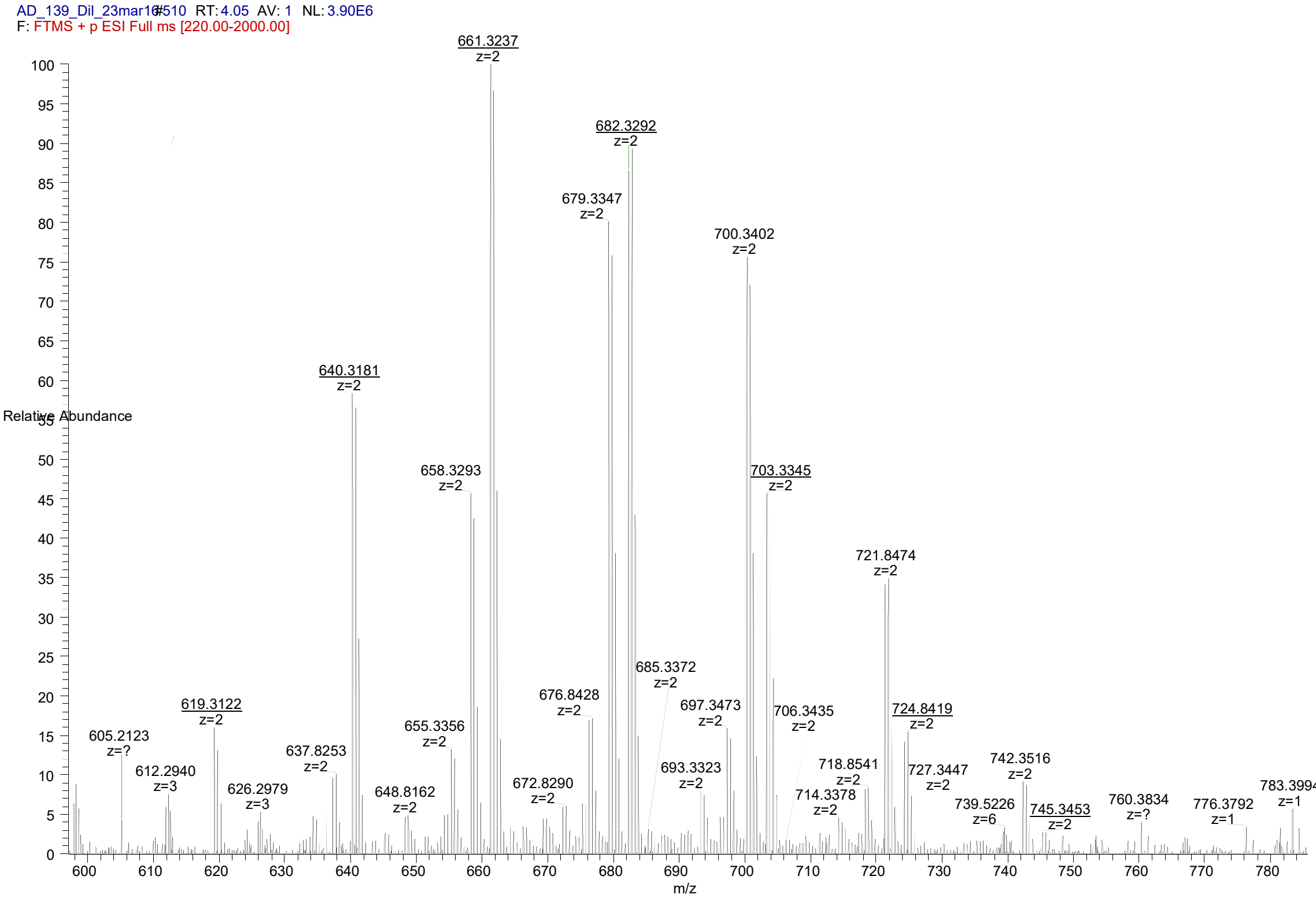

**Figure S3.** HR-Mass spectrum of sample **1s** in the 600-800 m/z range. The underlined m/z refer to the tetramer in the various amide shapes.

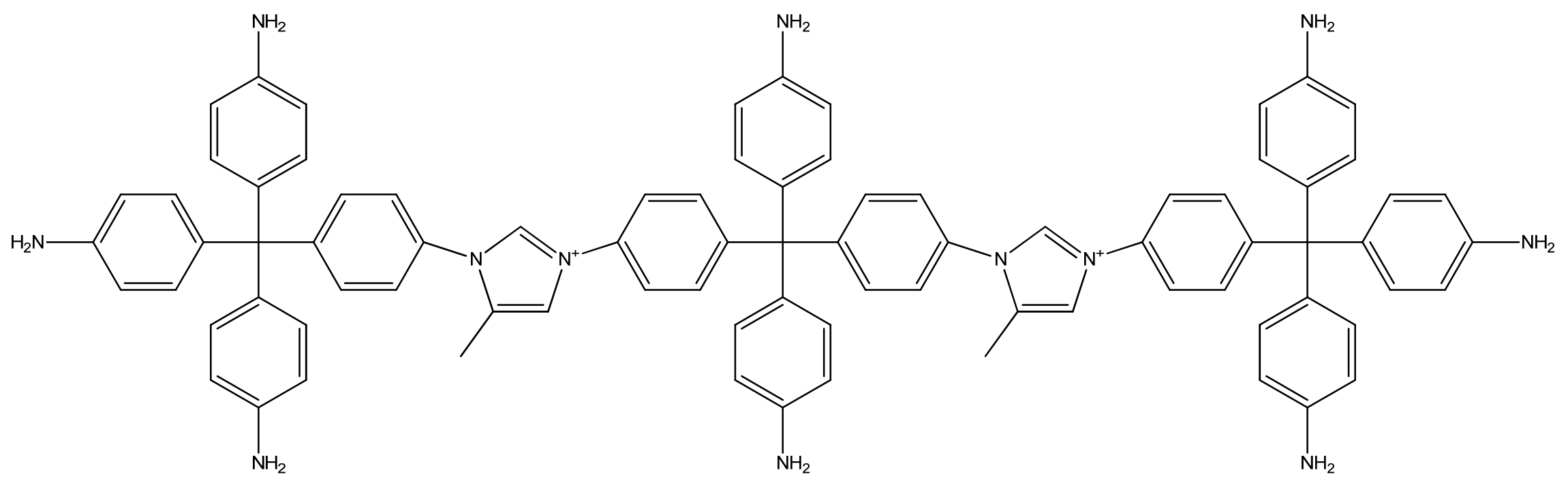

### 3.4 *Tetramer identification from the HR mass spectreometry*

In the mass spectrum reported in Figure S4 the signals related to the tetramer and to the tetramer with various amides on the free amines are present.

Table S4 reports the exact m/z ratio of the tetramer and of the tetramer with growing number of amides. The molecules evident in the mass spectra reported in Figure S4, are only the tetramers with one to five of the amino groups in form of amides, because only these molecules are statistically favoured at the used concentration of acetic acid in solution giving a signal high enough to be detected by the mass spectrometer.

**Table S4.** Exact m/z ratio of the tetramer, and of the tetramer with growing number of amides.

| number of acetamide | m/z |
|---|---|
| 0 | 556.2752 |
| 1 | 570.2787 |
| 2 | 584.2822 |
| 3 | 598.2858 |
| 4 | 612.2893 |
| 5 | 626.2928 |
| 6 | 640.2963 |
| 7 | 654.2999 |
| 8 | 668.3034 |
| 9 | 682.3096 |
| 10 | 696.3104 |

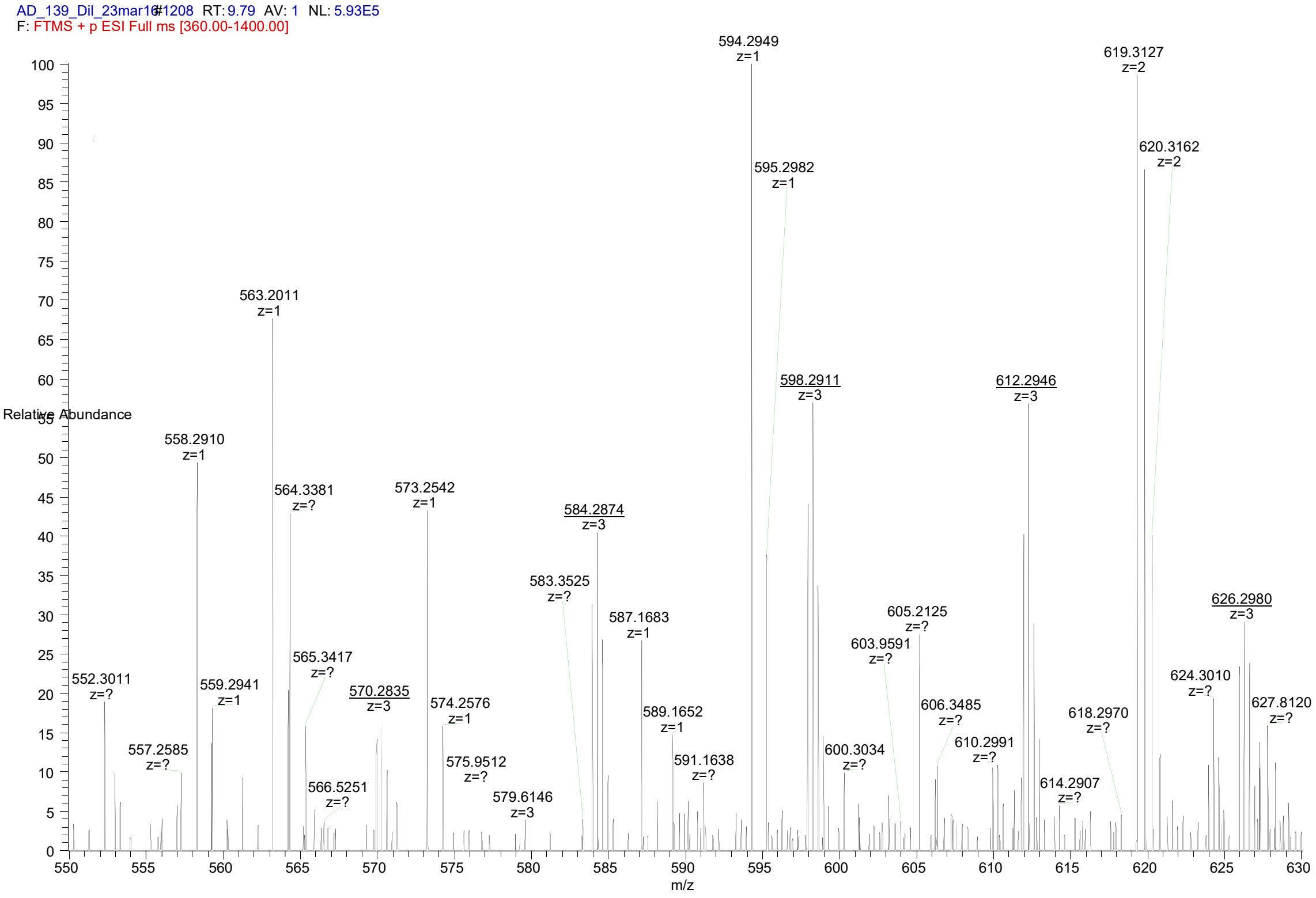

**Figure S4.** HR-Mass spectrum of sample **1s** in the 550-630 m/z range. The underlined m/z refer to the tetramer in the various amide shapes.

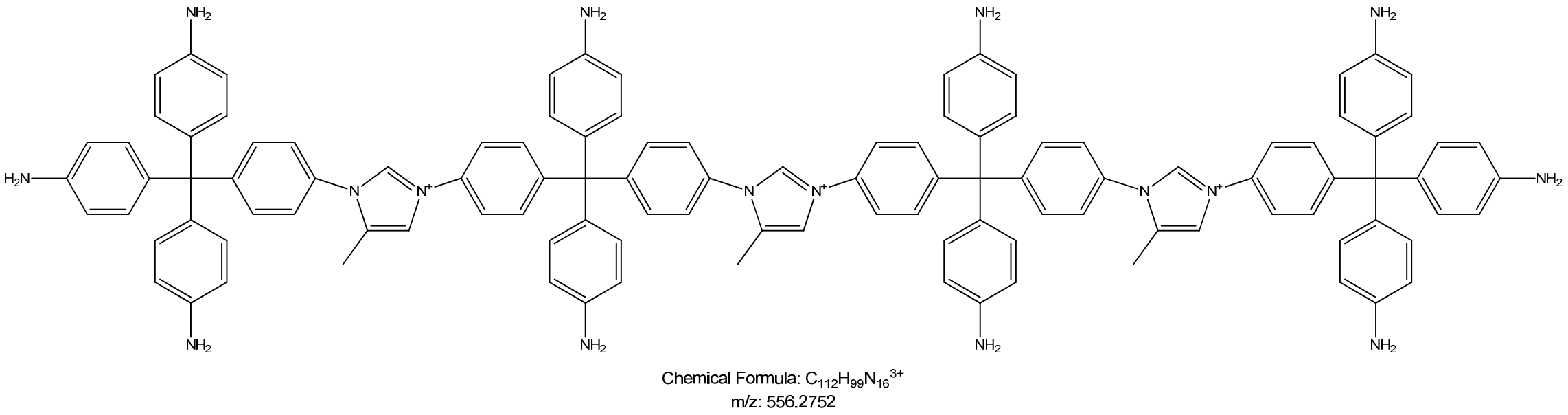

### 3.5 MS$^n$ Fragmentation of the pure dimer used to prove the chemical structure

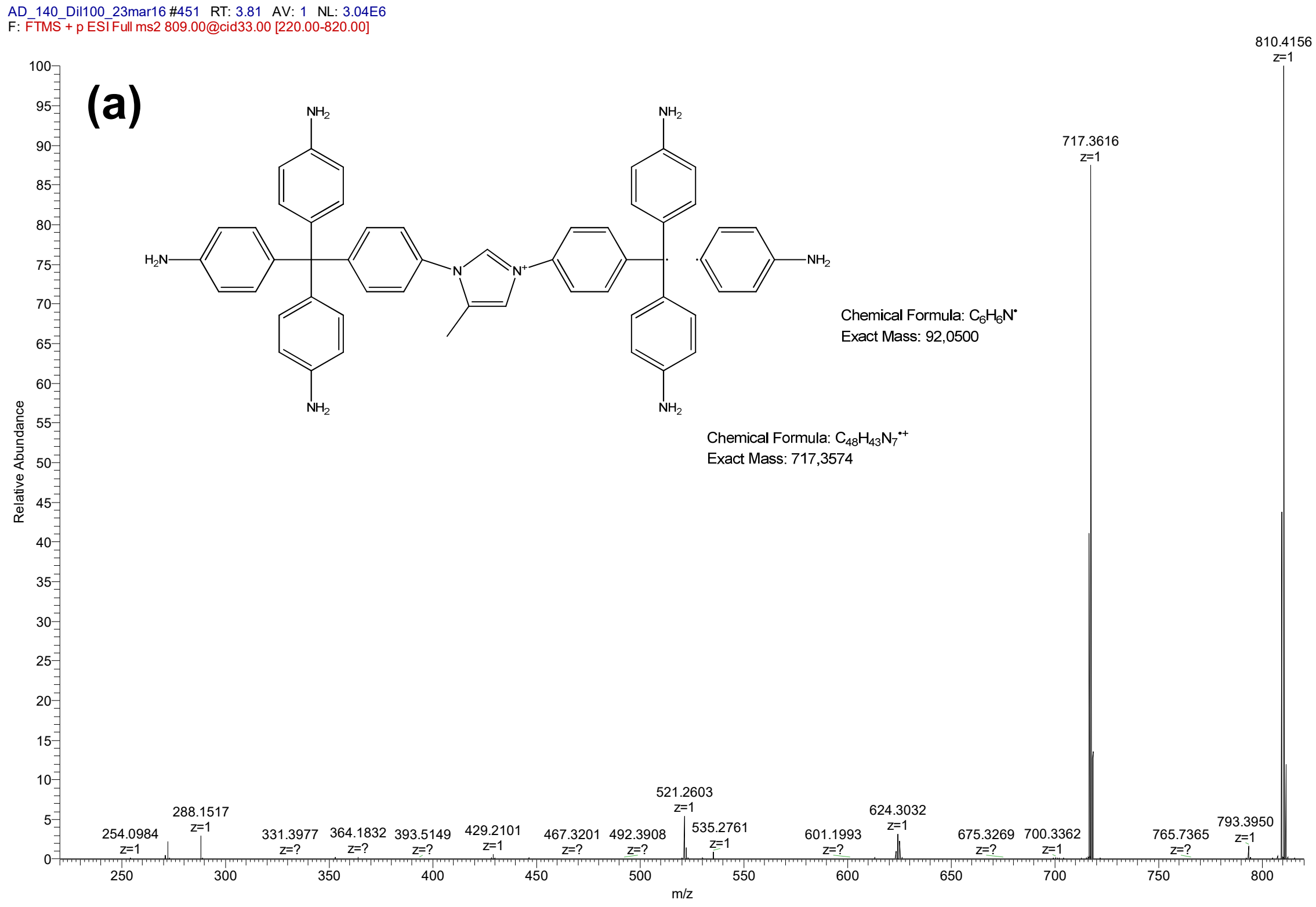

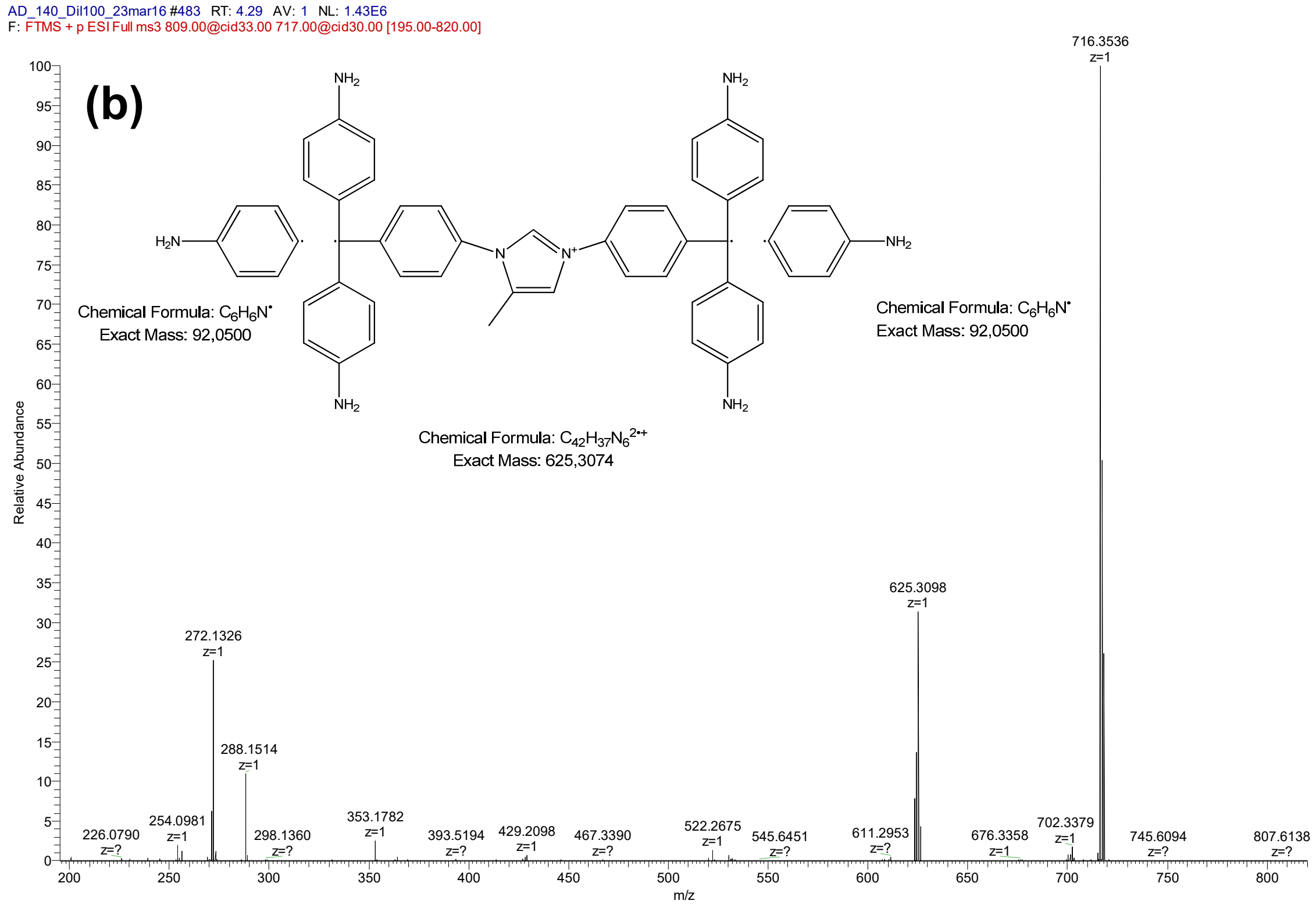

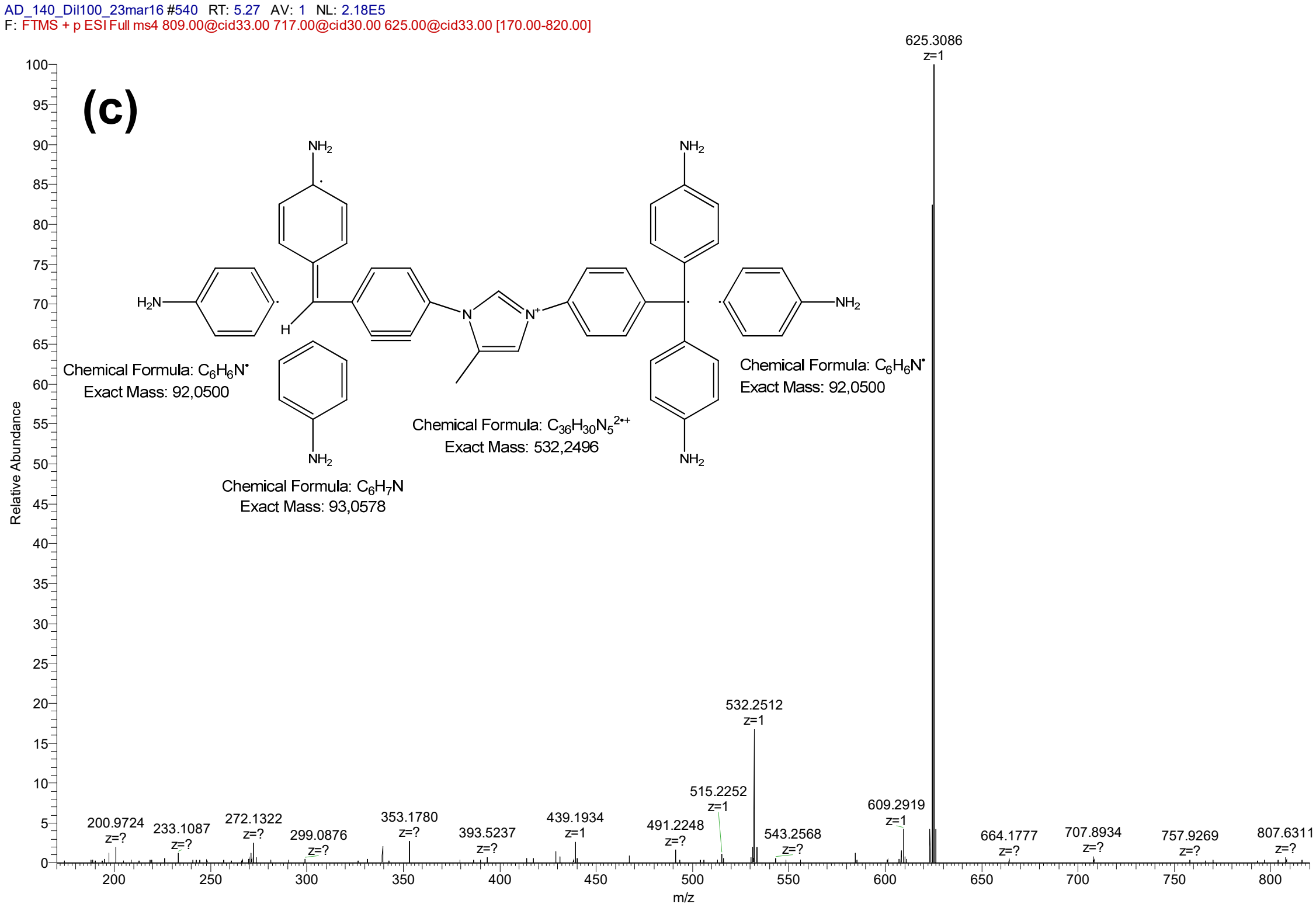

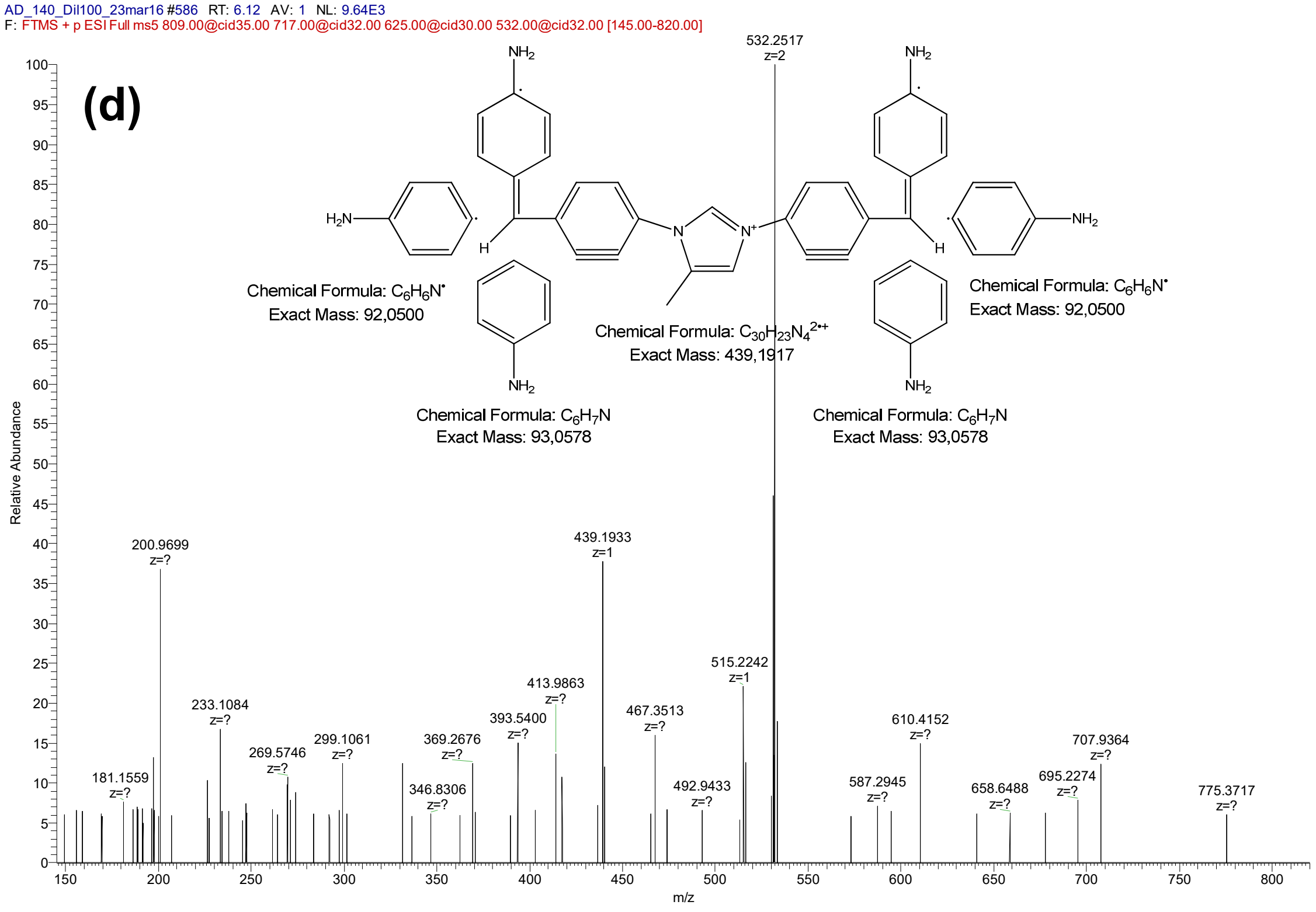

**Figure S5.** $MS^n$ fragmentation of the dimer without amide moiety in the 150-800 m/z range. Part a refers to $MS^2$, part b to $MS^3$, part c to $MS^4$, part d to $MS^5$. Every MS spectrum has a proposed fragmentation scheme.

## 3.6 MS$^n$ Fragmentation of the dimer with one acetamide in order to prove the chemical structure

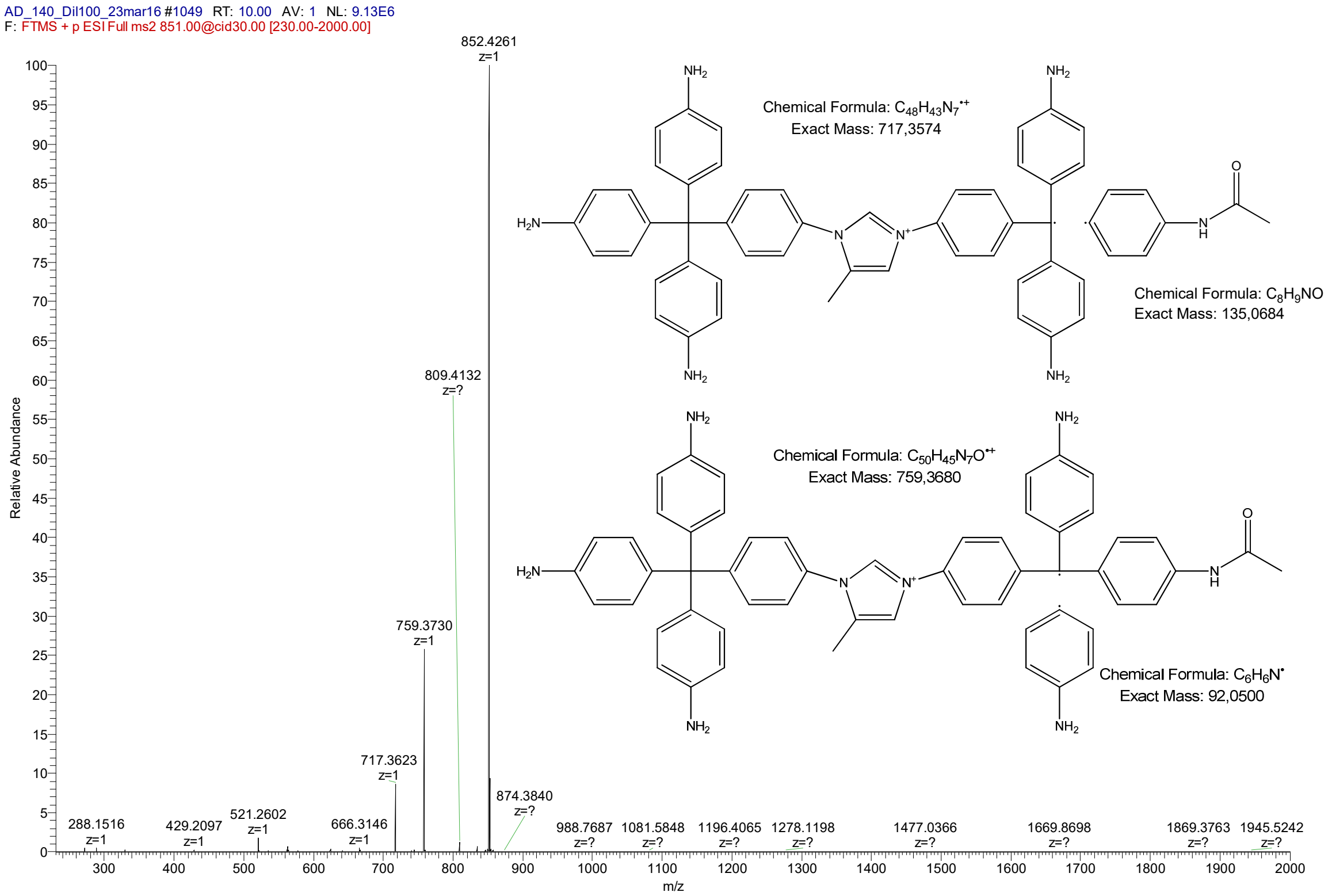

**Figure S6.** MS$^2$ fragmentation of the dimer with one amide in the 150-800 m/z range. Two fragmentation pathways occur together due to the neutral loss of unit with and without amide; both have a proposed fragmentation scheme.

## 3.7 $N_2$ adsorption isotherms at 77K

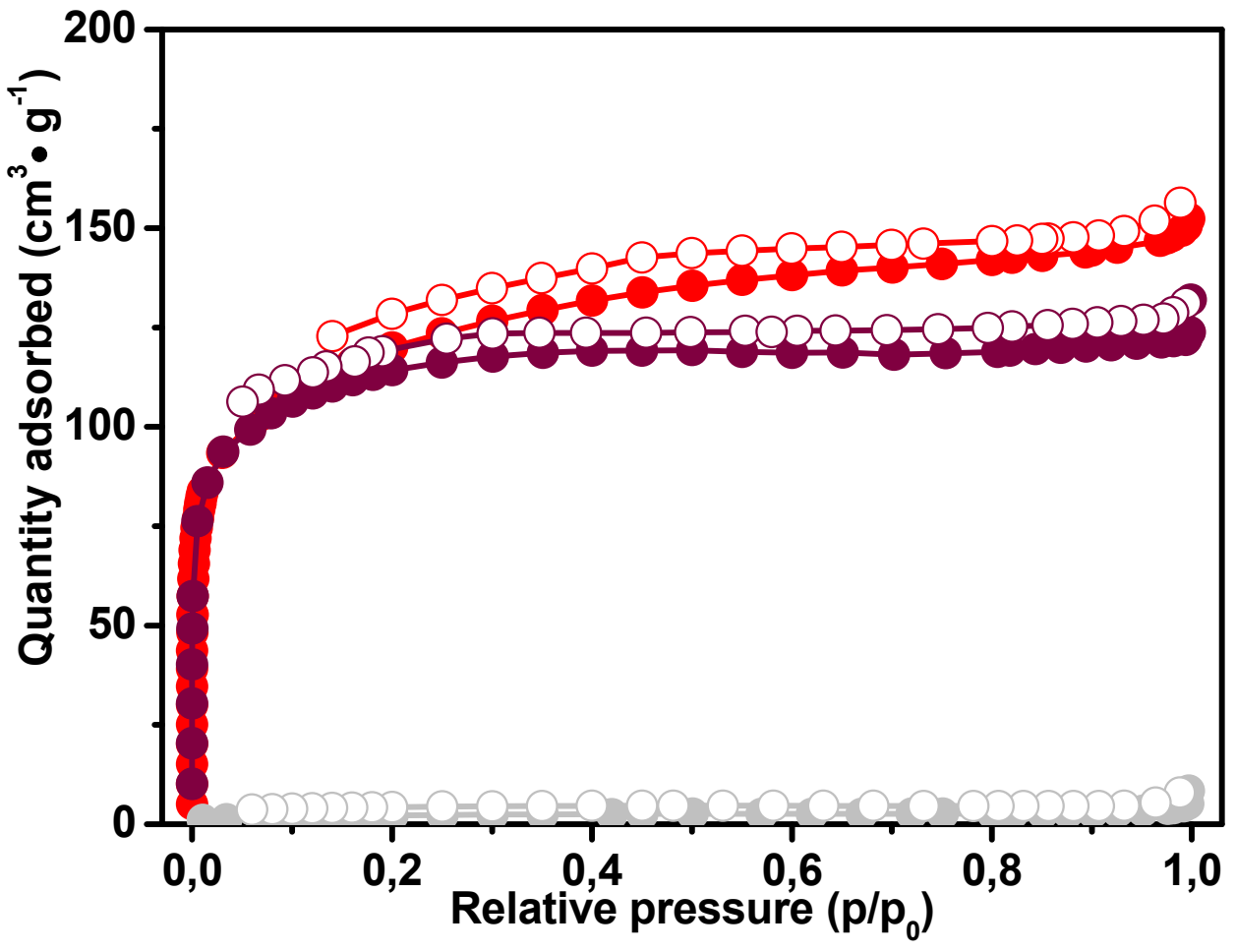

**Figure S7.** $N_2$ adsorption isotherms at 77K for samples: **1** (red curve), **1a** (rusty curve), **1b** (light gray curve).

### *3.8 Pore size distribution*

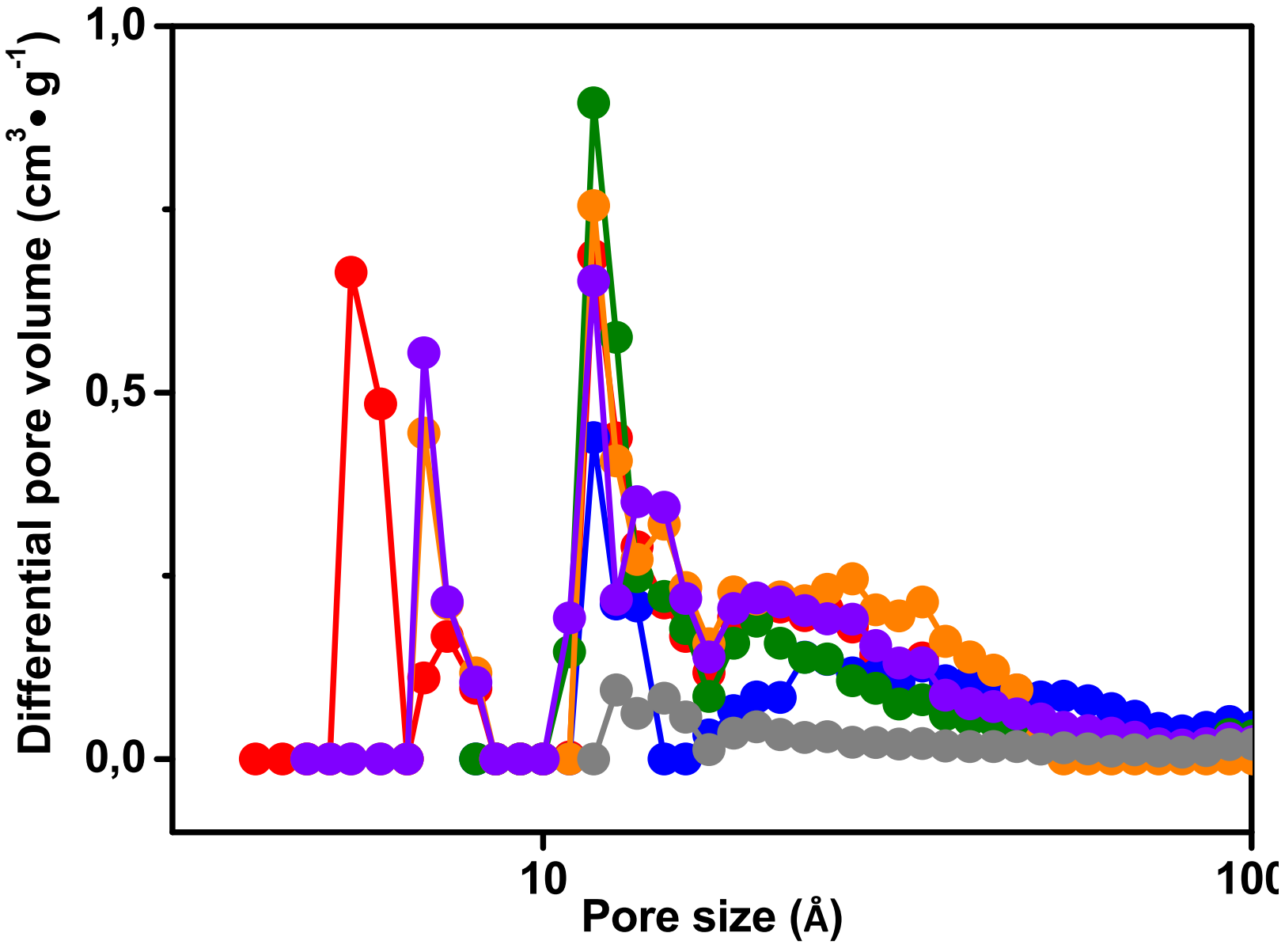

**Figure S8.** Pore size distribution obtained from the $N_2$ adsorption isotherms at 77K using non-local density functional theory (NL-DFT) and pore model for carbon with slit pore geometry. Results of the isotherms analysis are reported for samples: **1** (red curve), **2** (blue curve), **3** (green curve), **4** (orange curve), **5** (violet curve) and **6** (dark grey curve).

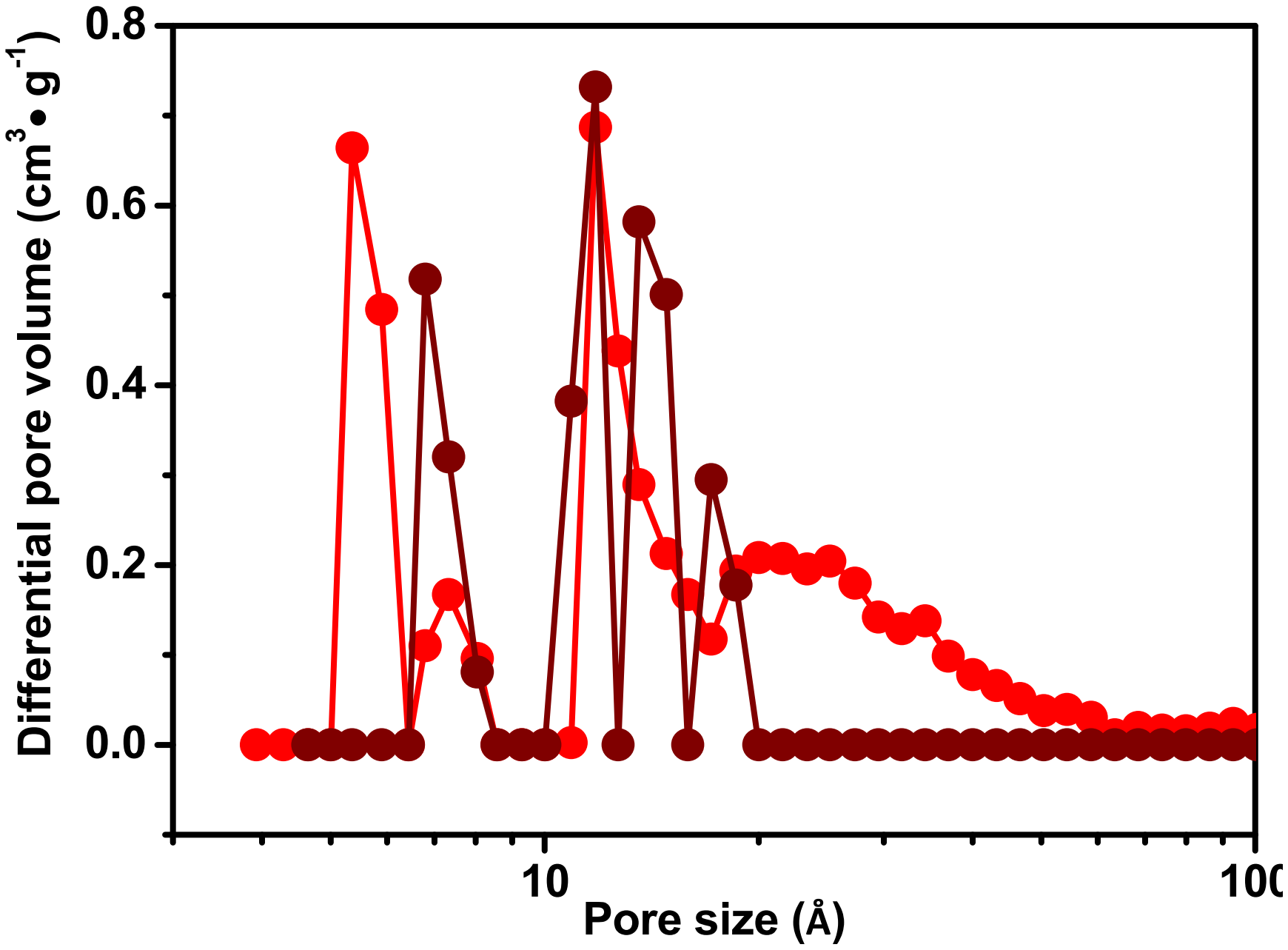

**Figure S9.** Pore size distribution obtained from the $N_2$ adsorption isotherms at 77K using non-local density functional theory (NL-DFT) and pore model for carbon with slit pore geometry. Results of the isotherms analysis are reported for samples: **1** (red curve), **1a** (rusty curve).

### *3.9 SEM image*

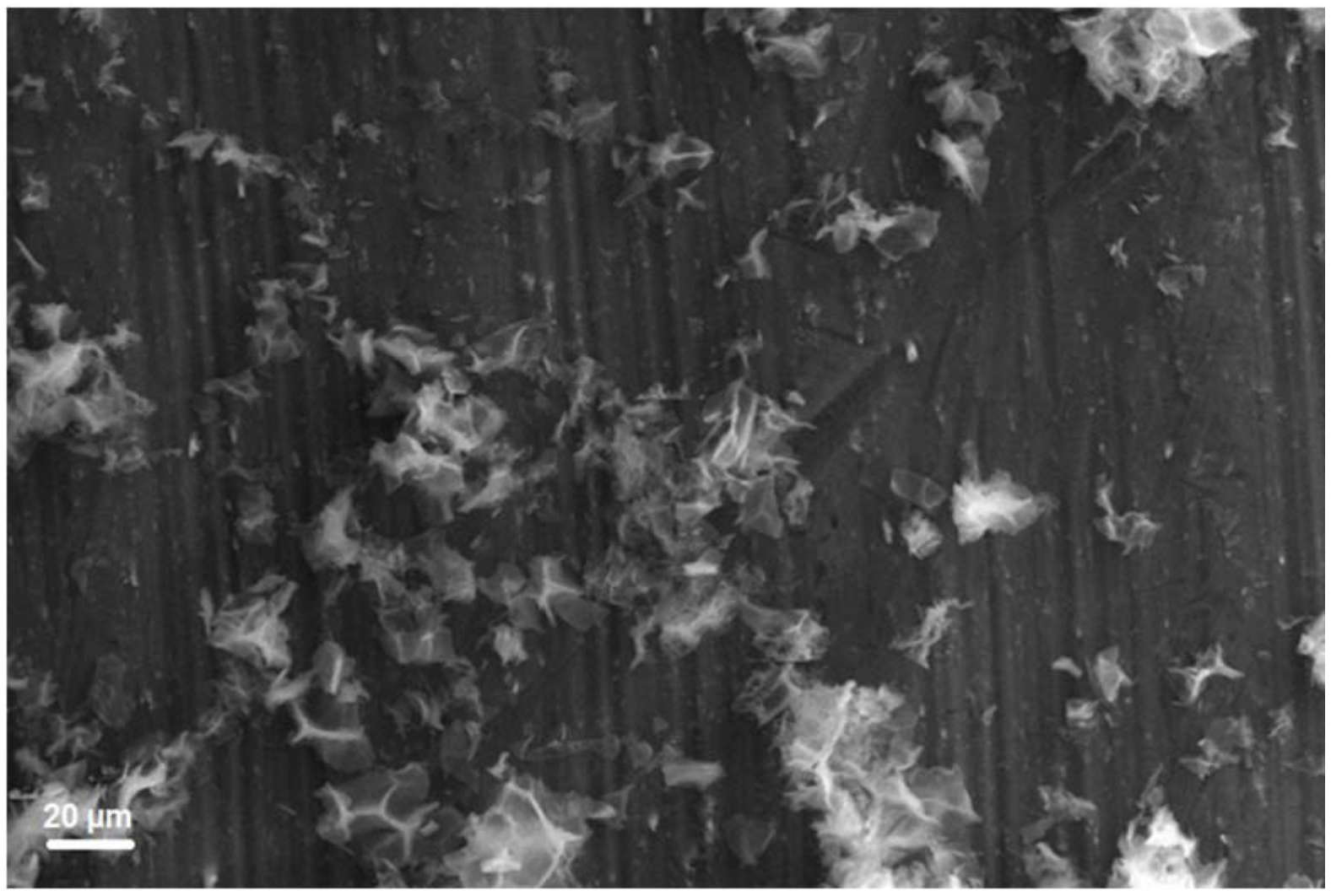

**Figure S10.** SEM picture of sample **1**

### *3.10 Thermogravimetric analysis*

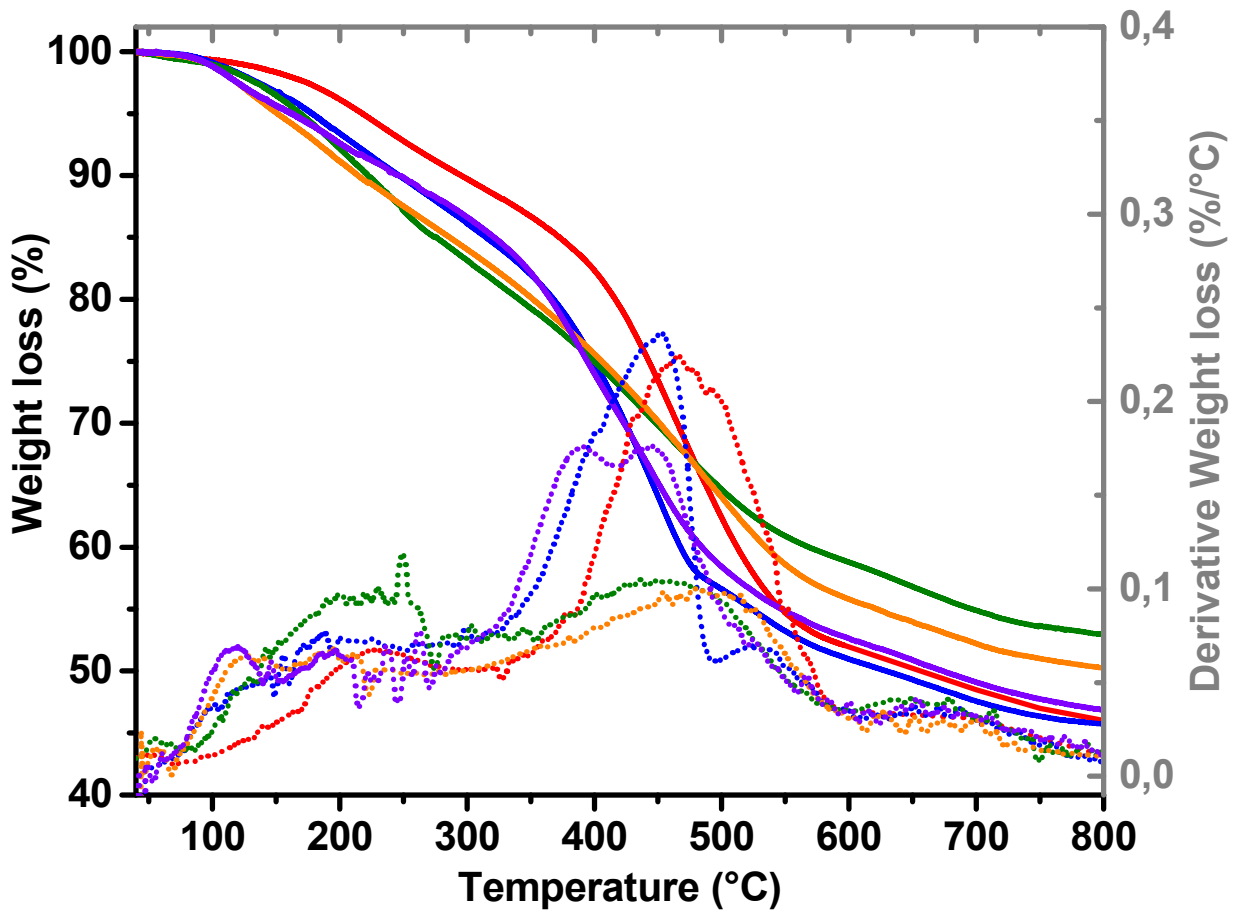

**Figure S11.** Thermogravimetric profiles (full curves) and derivative weight loss (dotted curves) for samples: **1** (red), **2** (blue), **3** (green), **4** (orange) and **5** (violet).

### *3.11 Carbon dioxide adsorption measurement*

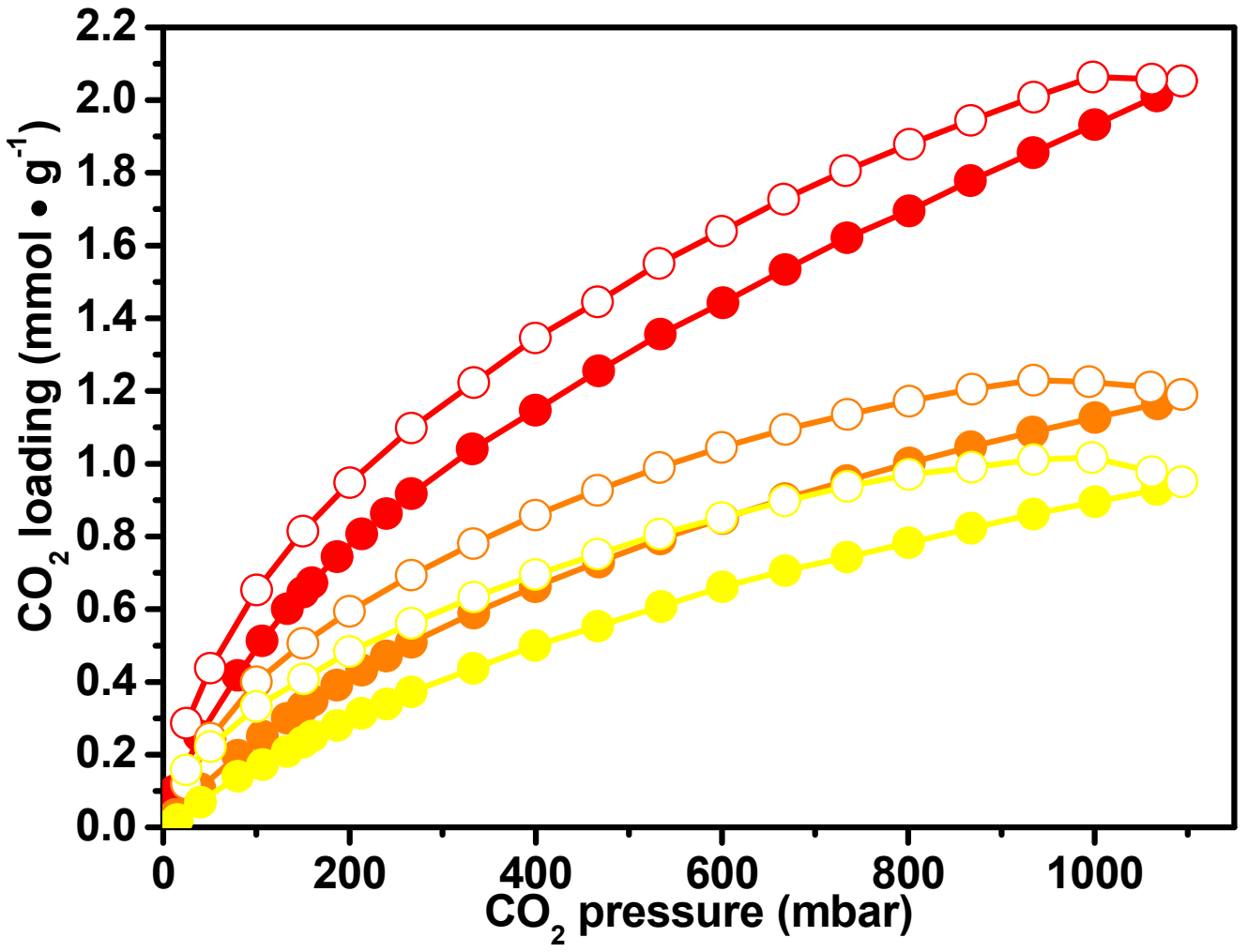

**Figure S12.** Carbon dioxide adsorption isotherm for samples **1** at: 273K (red curve), 298K (orange curve), 313K (yellow curve). Spheres and circles describe the adsorption branch and the desorption branch respectively.

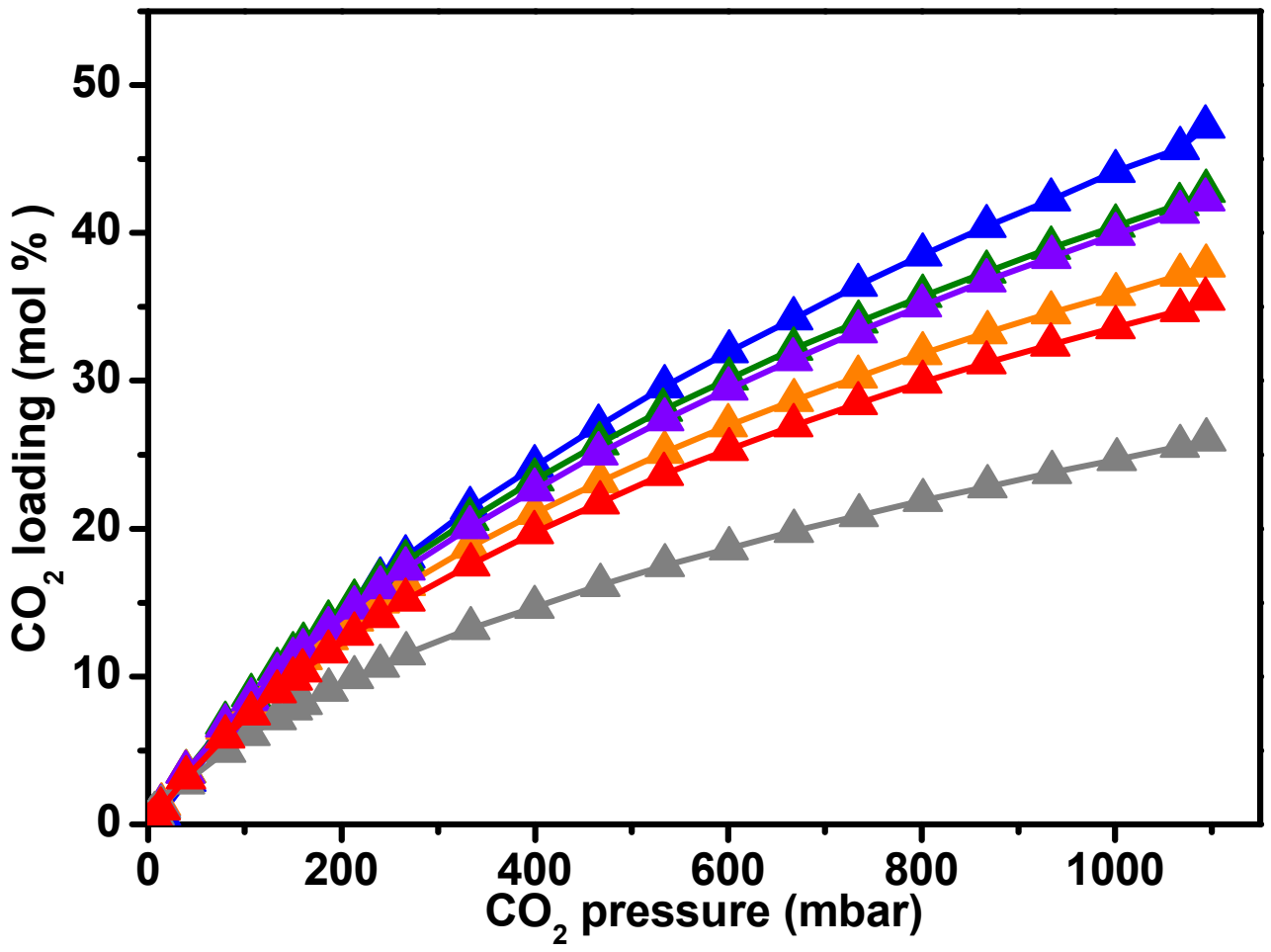

**Figure S13.** Carbon dioxide adsorption isotherms represented in mol %. of $CO_2$ respect to the imidazolium at 298 K for samples: **1** (red curve), **2** (blue curve), **3** (green curve), **4** (orange curve), **5** (violet curve), **6** (grey curve). The desorption isotherms are not reported for sake of view.

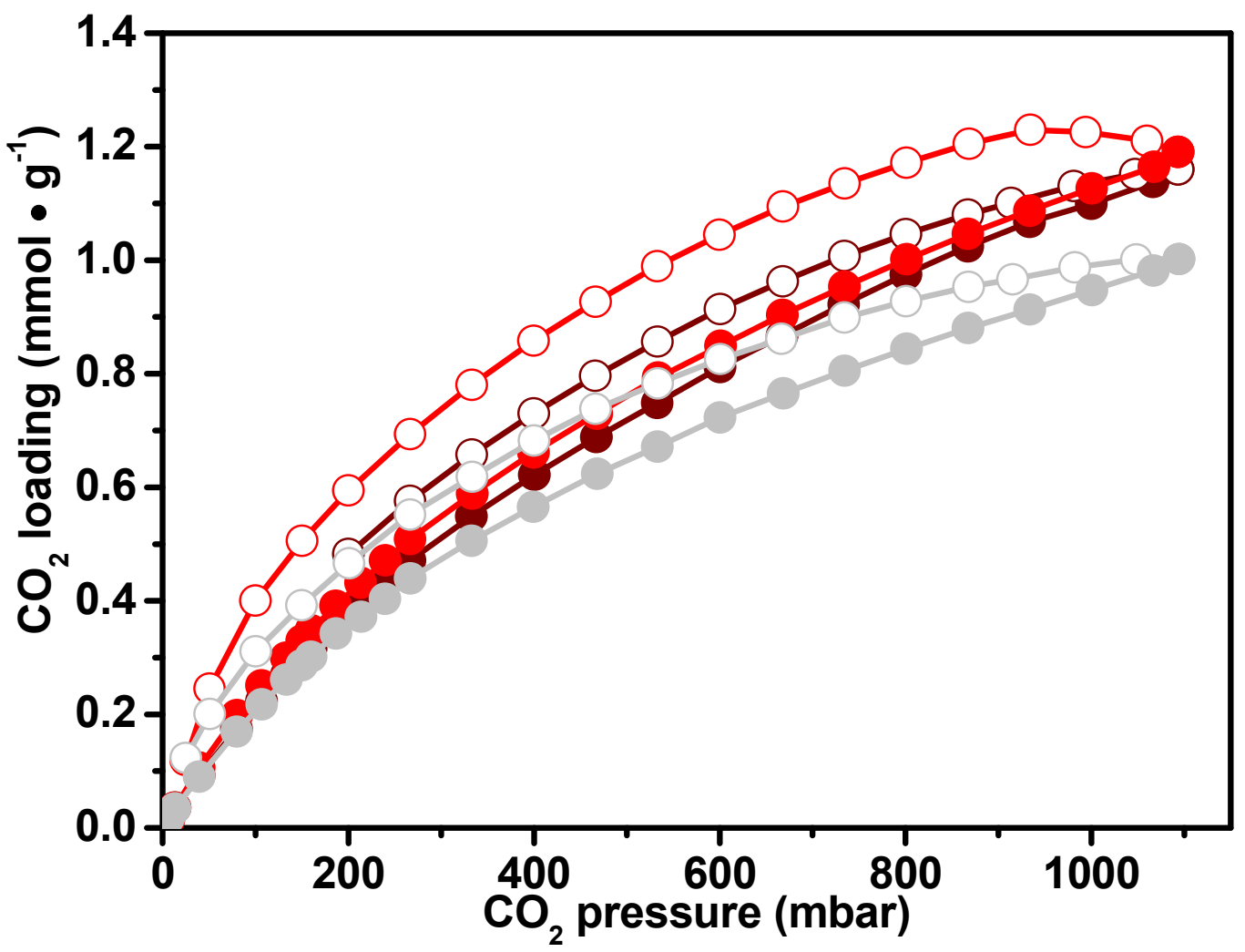

**Figure S14.** Carbon dioxide adsorption isotherm at 298K for samples: **1** (red curve), **1a** (rusty curve), **1b** (light gray curve). Spheres and circles describe the adsorption branch and the desorption branch respectively.

### *3.12 Micro-calorimetry*

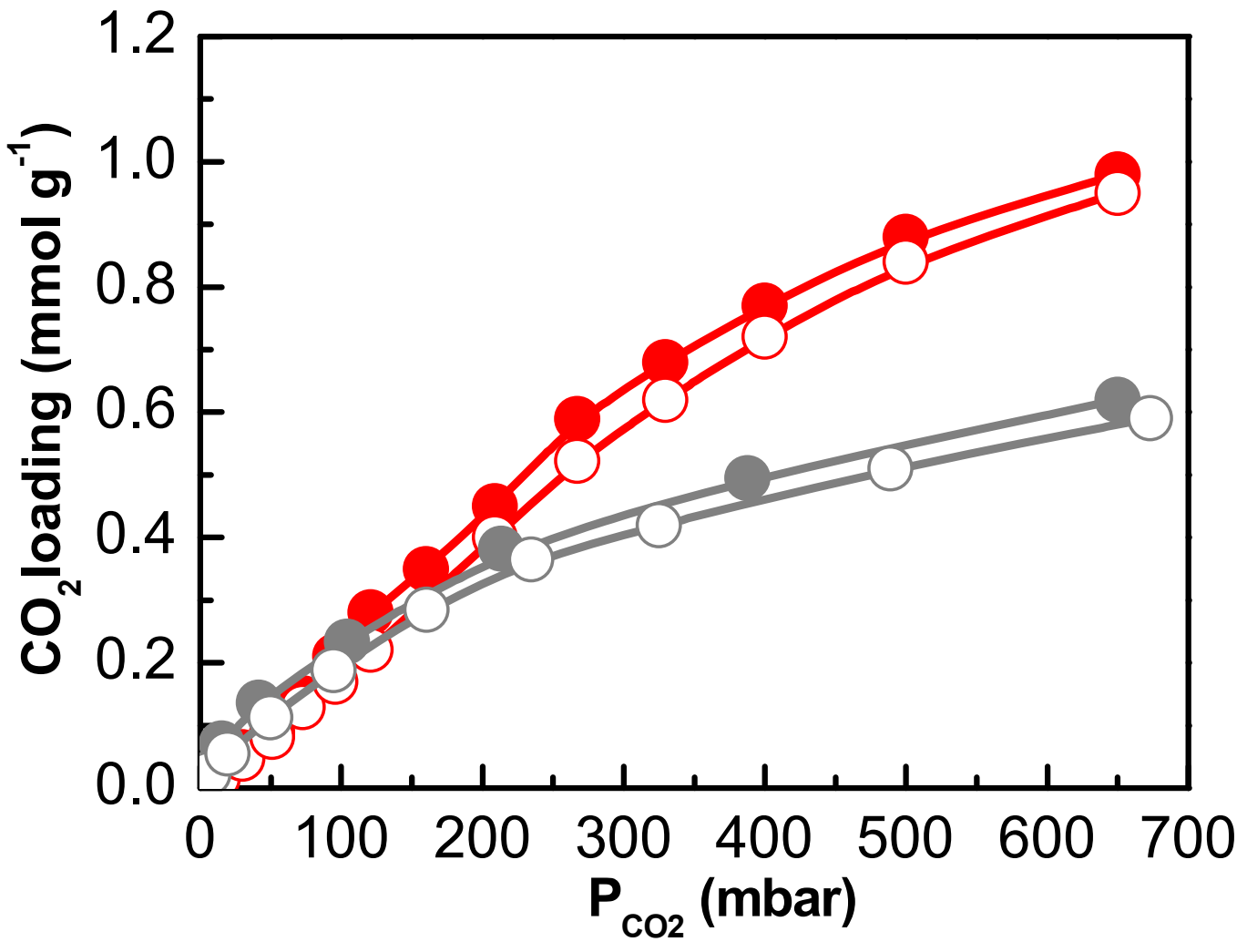

**Figure S15.** Adsorbed amount as a function of the equilibrium pressure relative to the adsorption at 298 K of $CO_2$ on samples **1** (red curves) and **6** (grey curves and dots).Spheres refer to the primary adsorption run, whereas circles refer to the secondary adsorption run.

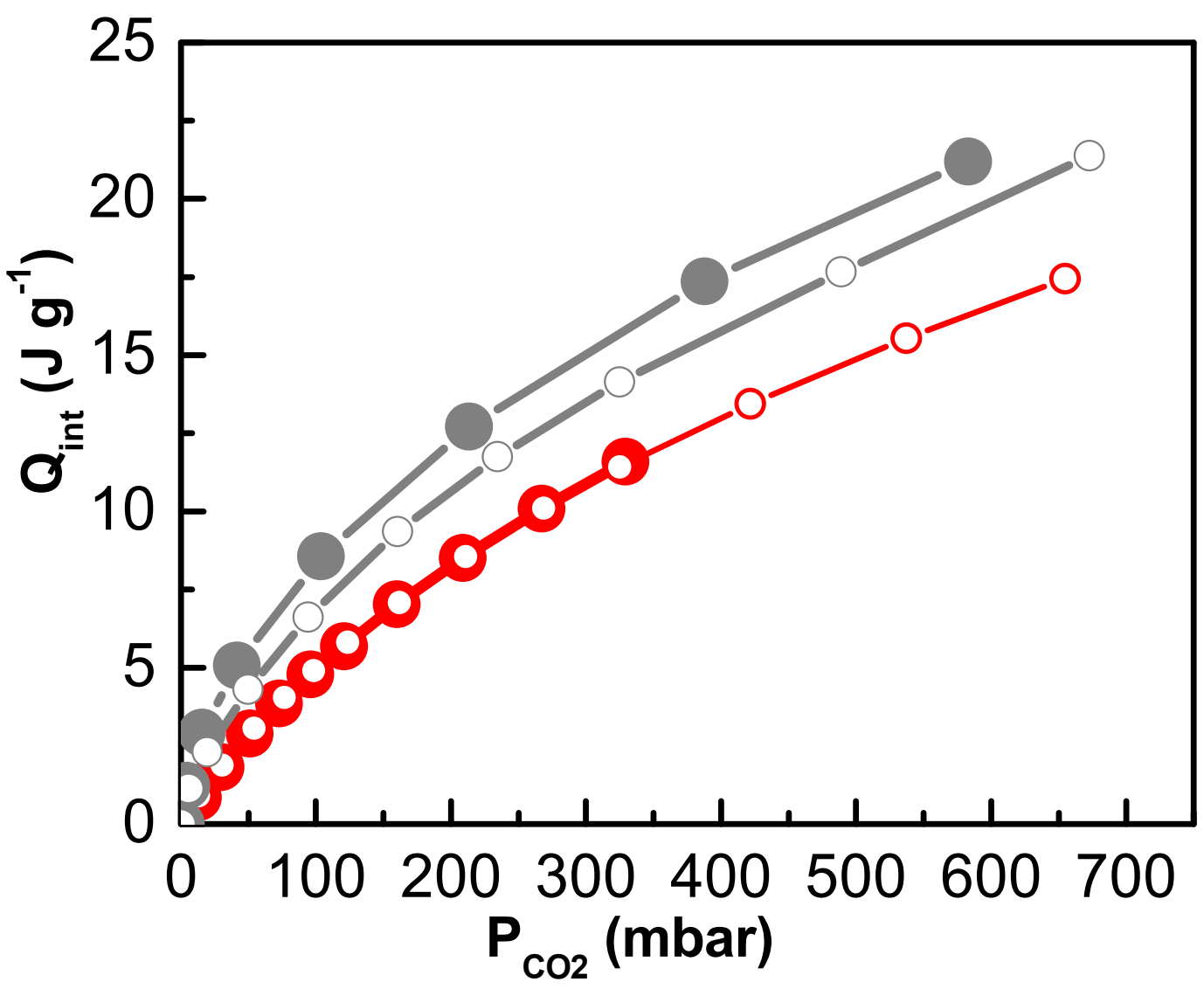

**Figure S16.** Integrated heat of carbon dioxide adsorption as a function of $CO_2$ pressure for samples **1** (red curve and dots) and **6** (grey curve and dots). Closed spheres refer to primary adsorption, while open circle refer to secondary one.